# The Phenomenological Scaling Relations of Disk Galaxies

Jeffrey M. La Fortune
1081 N. Lake St., Neenah, WI USA 54956 forch2@gmail.com
16 December 2019

*Abstract*

A conservative scaling model based on Newton's law of circular motion and the virial theorem is proposed. Employing dimensionless scaling relations, a physical solution to the Baryonic Tully-Fisher Relation (BTFR) is obtained and the origin of the Radial Acceleration Relation (RAR) is provided. A simple model of the Milky Way's dynamics reveals excellent agreement with recent Gaia DR2 kinematical results.

*Introduction*

Scaling-homology is a principle that infers physical self-similarity in a particular class of objects. In the 'extreme' interpretation these objects are considered "exact scaled copies" of one another. As a result, galactic physical properties respect strict relations between scale and proportionality between them (Novak 2012). These homological scaling relations (power laws) have been known for over a decade and have been extensively explored (Zaritsky 2008). This proposal extends the concept of scaling-homology using the latest astrometric data available. We employ the Milky Way (MW) as the scaling exemplar and extend its scaling relations to a wider class of rotationally supported galaxies inhabiting the local universe.

This scaling investigation imposes two well established physical constraints. The first is that disks obey the spin parameter equation $\lambda = JE^{1/2}/GM_{Dyn}^{5/2}$ with dynamic mass ($M_{Dyn}$), angular momentum (J) and total energy (E) (Peebles 1971). The dimensionless spin parameter ($\lambda$) is treated as a constant in agreement to the empirically determined value 0.423 (±0.014) (Marr 2015b). Proportionality between dynamic and baryonic mass is determined by the dimensionless mass discrepancy $\mathcal{D} = M_{Dyn}/M_{Bar}$ or its equivalent $(V_{Obs}/V_{Bar})^2$. For all galaxies, mass discrepancy is fixed to $\mathcal{D}=5.9$ with this value conforming to an average baryonic fraction $f_b = 0.17$, although the model allows for any reasonable value. These strict constraints maintain a physical consistency between scaling relations, independent of individual galactic mass and morphology.

In this approach we define a small number of parameters, the first being $R_{Disk}$ ($R_D$). This is the terminal disk radius indicating the outer edge of HI gas (typically $\Sigma_{HI} \leq 1 M_\odot pc^{-2}$). The next is observed circular velocity ($V_{Circ}$). Galactic dynamic mass ($M_{Dyn}$) is obtained via Newton's law for circular motion $M_{Dyn} = R_D V_C^2/G$. In this regard, $M_{Dyn}$ is equivalent to total galactic mass or in as treated in ΛCDM, dark matter halo mass ($M_{DM}$). This global approach reduces model complexity and permits all disk galaxies to be defined by three rigidly defined and physically measurable parameters, independent of motivation (modified gravity, dark matter, or otherwise).

Another crucial scaling constraint is the thermodynamic nature and properties of disk galaxies. In this regard, all galaxies are treated as self-gravitating "open systems" existing in quasi-equilibrium with local "surroundings." Thus, the kinematical properties at the virial boundary must be described by a statistical probability distribution. To describe stellar kinematics at the virial surface, a conventional Maxwell-Boltzmann velocity distribution is employed. We find this distribution explains issues experienced pinning down a precise total mass estimate of the Galaxy.

This paper consists of a short main body where scaling parameters defined and measured for the MW and M31. These co-exemplars establish basic scaling principles which then can be extended to the wider class of disk galaxies under a universal scaling relation, termed [$M_{Dyn}$-$R_{Disk}$-$V_{Circ}$]. This work relies heavily on rotation curves which accurately trace the galactic potential from all constituents as an aggregate.

We relate the Baryonic Tully-Fisher Relation (BTFR) to scaling expectations and make comparisons with two popular competing paradigms, ΛCDM and MOdified Newtonian Dynamics (MOND). We also explore the physical nature of the Radial Acceleration Relation (RAR) (Milgrom 1983) (McGaugh 2016). Previous work with the Burkert dark matter halo profile is examined in detail as it shares similar characteristics to the observed scaling parameters (Burkert 1995). Using updated data, we investigate the galactic dynamic surface density-acceleration dependence and connection with scaling results (Gentile 2009).

*Milky Way (MW) - Observed Scaling Parameters*
We establish scaling dimensions, parameters, and ratios employing the 'grand rotation curve' obtained for the MW (Sofue, Dark Halos of M31 and the Milky Way 2015). In Figure 1 below, we augment Sofue's original profile with data including all rotation *and* velocity dispersion components (King III 2015). This is a new method of obtaining salient parameters and scaling relations when the dispersion is interpreted as true thermodynamic phenomena. The figure provides a simple but powerful version of galactic dynamics that are based on observations and application of classical law. $V_{Peak}$ ($V_P$) is the maximum velocity obtained from the three component dispersions observed at 23 kpc ($R_P$) from King III.

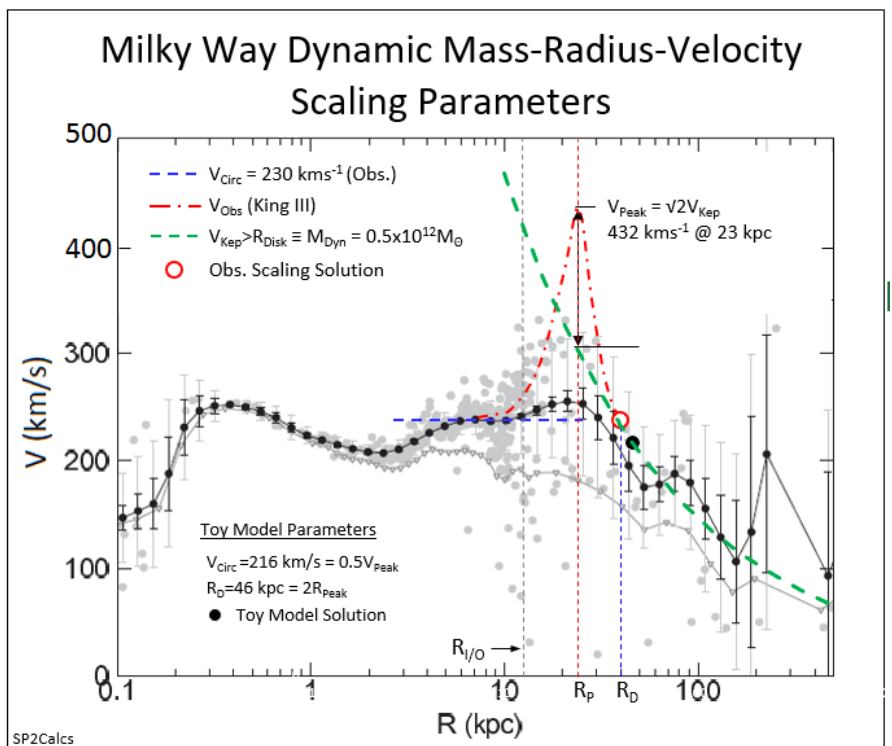

*Figure 1: The MW rotation curve with circular velocity (horizontal blue solid), the Keplerian $M_{Dyn}$ fit (green dash), dispersion velocity data (red dot-dash) and*

*inner/outer disk radius (vertical dark gray dash) with the intersection denoted by the open red circle. The dispersion driven velocity peak (red dot-dash) is from Fig. 10* (King III 2015).*The scaling parameters for toy model are given in the key-lower left. Image source - Fig.3* (Sofue 2015)

From above figure we obtain the MW specific scaling values for $M_{Dyn}$, $R_{Disk}$ and $V_{Circ}$:
- Total dynamic mass $M_{Dyn}$ = 0.5 x $10^{12} M_\odot$ as a central with Keplerian velocity >$R_D$
- Outer disk radius $R_D$ = 40 kpc (HI gas disk edge, not exponential scale length)
- Rotation velocity $V_C$ = 230 kms$^{-1}$ (Pato 2015)

The MW dimensionless scaling ratios and peak velocity:
- Inner/outer disk radius ratio: $R_{I/O}$=0.5$R_P$=11.5 kpc
- Virial Velocity (peak velocity) $V_{Peak}$ = 432 kms$^{-1}$ = √2$V_{Kep}$ at $R_{Peak}$ = 23 kpc
- Peak velocity/outer disk radius ratio: 23 kpc/40 kpc = 0.58

With above scaling parameters, we find the MW obeys Newton's law for circular motion:

$$M_{Dyn} = R V^2/G = 40\ kpc * (230\ kms^{-1})^2/4.3\ x\ 10^{-6}\ kpc M_\odot^{-1} km^2 s^{-2} = 0.5\ x\ 10^{12} M_\odot$$

Equating MW's local peak velocity to the local escape velocity at $R_P$:

$$V_{Esc} = \sqrt{2\ GM_{Dyn}/R} = \sqrt{2*0.5x10^{12}*4.3\ x10^{-6}/23} = 432\ kms^{-1}$$

Observed scaling parameter values for the MW intersect at the red open circle. A recent article stresses the importance of including both rotational support and velocity dispersion components to obtain physically realistic galactic dynamic mass estimates (Turner 2017). Appendix C provides observations that include both rotation and dispersion support extensively utilized in this analysis (King III 2015). To appreciate the physical response of the Galaxy's virial/Newtonian mass duality, we reproduce a compilation of satellite dynamics out to 400 kpc in Appendix D including "best fit" ΛCDM galactic Model III (Bajkova 2017). A full discussion related to this data set is included in the Appendix.

In later sections, we equate $V_P$ = $V_{Esc}$ as the Maxwell-Boltzmann "most probable" velocity ($V_{MP}$) due to the statistical nature of galactic velocity dispersion. Figure 1 also includes the toy model normalized scaling parameters 2$R_P$=$R_D$=46 kpc and $V_C$=216 kms$^{-1}$ (black dot) detailed in Appendix E illustrating the contributions of disk E and J as a function of radius and mass discrepancy.

*Andromeda Galaxy (M31) - Observed and Modeled Scaling Parameters*
Sofue's extended rotation profile for M31 is shown in Figure 2. As with the MW, M31's total galactic mass is estimated at $M_{Dyn}$ = 1.47 x $10^{12} M_\odot$. This agrees with the Sofue estimate $M_{DM}$≈1.4 - 2.0 x $10^{12} M_\odot$ using halo best fit parameters. Similar to the MW, just beyond the outer disk, the rotation profile follows a conventional Keplerian decline (green dash). As there are no component dispersions available for M31, the same scaling treatment is used to determine $V_P$. An important point to emphasis is that beyond $R_P$ for both the MW and M31, uncertainty and velocity distribution range is significantly higher than inside $R_P$. This is a key indicator of the presence of a virial partition within the disk at $R_P$, the virial boundary.

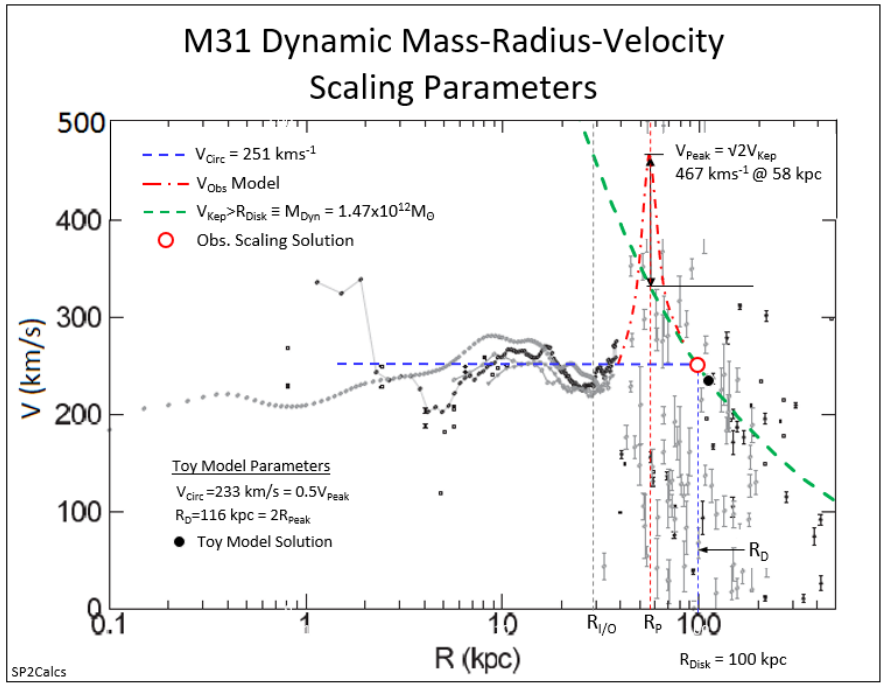

*Figure 2: M31 rotation curve - same format to Figure 1 with scaling relations based on MW parent model. Image source - Fig.1* (Sofue 2015)

M31's observed rotation curve provides the following:
- Total dynamic mass $M_{Dyn} = 1.47 \times 10^{12} M_\odot$ as a central mass with Keplerian velocity $>R_D$
- Outer disk radius $R_{Disk} = 100$ kpc
- Rotation velocity $V_{Circ} = 251$ kms$^{-1}$

M31 follows simple size proportionality (as a scaled copy of the MW):
- Inner/outer disk radius ratio: $R_{I/O} = 0.5 R_P = 29$ kpc
- Virial Velocity (peak velocity) $V_{Peak} = 467$ kms$^{-1}$ = $\sqrt{2} V_{Kep}$ (330 kms$^{-1}$) at $R_{Peak}$
- Radius of peak tangential velocity to disk radius: $0.58 \times 100$ kpc = 58 kpc.
  (Note that 0.58 is a MW specific observation and M31's ratio may differ.)

The Keplerian determined dynamic mass is consistent with Newtonian dynamics:

$$M_{Dyn} = R V^2 / G = 100 \; kpc * (251 \; kms^{-1})^2 / 4.3 \times 10^{-6} \; kpc M_\odot^{-1} km^2 s^{-2} = 1.47 \times 10^{12} M_\odot$$

M31's peak velocity is obtained:

$$V_{Esc} = \sqrt{2 \, GM_{Dyn}/R} = \sqrt{2 * 1.47 \times 10^{12} * 4.3 \times 10^{-6} / 58} = 467 \; kms^{-1}$$

In Appendix F, we show M31's observed globular cluster radial density and velocity profile to illustrate the accuracy and usefulness of the scaling parameters in describing the galactic dynamic.

*Milky Way Dark Matter Halo and Scaling Parameter Comparison*

Early-era ΛCDM simulations incorporated rotation curve data nearer the solar radius and extrapolated well beyond the physical disk to satisfy its cosmological constraints. At that time, flat rotation was thought to extend to 250 to 300 kpc, the purported virial radius of the dark matter halo leading to highly inflated galactic total mass. With the advent of precision astrometrics to the edge of the Galaxy, recent data (as Sofue above) directly contradicts this supposition. For the scaling model, we use this recent data to accurately obtain the dynamic mass of the Milky Way.

Current dark matter halo models now employ fairly rigid physical constraints. A recent suite of simulations has been published consistent with observed local values for total matter density and the magnitude of vertical gravitational force (Bobylev 2017). These recent simulations are consistent with precision kinematics to 200 kpc, highly constraining dark matter halo properties, especially at outer radii. One Bobylev simulation 'Model VI' is shown in Figure 3 below. In this model, a Miyamoto-Nagai potential is used for the baryonic bulge and disk and a Hernquist potential for the dark matter halo. The dark matter halo potentials smoothly decline in monotonic fashion, exhibiting no significant substructure, a hallmark of the dark matter halo density contrast.

Figure 3 also contrasts Model IV against the King total component velocity dispersion peak. The figure identifies a key issue - using incomplete rotation curves to construct a global halo model. We see that in the region of high dispersion, all data is missing (shaded region) and this missing data holds the key to understanding the physical nature of the Galactic dynamic, including the origin of mass discrepancy or 'missing mass.'

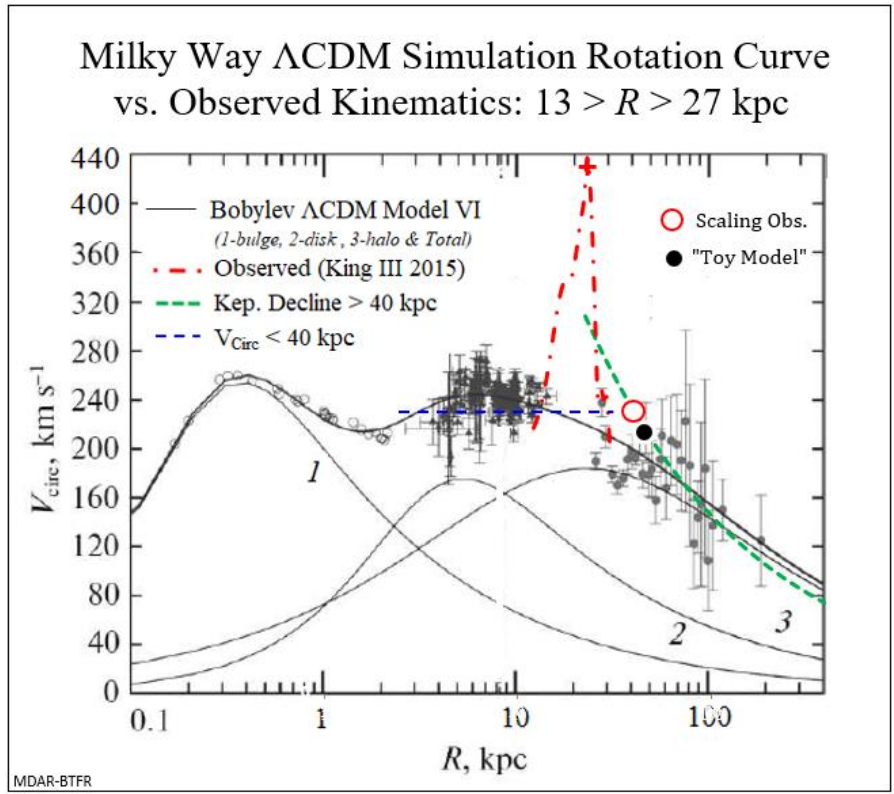

*Figure 3: MW total rotation curve data and Model VI, which is the best fit ΛCDM simulation. The velocity contributions are: (1-bulge, 2-disk, and 3-dark matter halo – all black solid curves). King III observed combined velocity dispersion components are superimposed (red dot-dash). Included is the observed 'R-V' scaling solution (open red circle) and the "Toy Model" (black point). Circular velocity is flat (blue dash) to disk edge where a Keplerian follows (green dash). Also indicated is the magnitude and location of peak $V_{MP}$=432 kms$^{-1}$ at $R_P$=23 kpc (red-cross). Image source - Fig.1 lower right panel* (Bobylev 2017)

With inclusion of accurate data in the Galaxy's outer regions, dark matter halo mass and extent have been drastically reduced over the years to fit the observed (and real) Keplerian decline beyond $R_D$. It is not a coincidence that Bobylev's dark matter halo properties now have maximal halo support at 21 kpc and $M_{DM}$≈0.49x10$^{12}$M$_\odot$, very similar to the scaling model. While disk E and J can conspire to physically produce the significant velocity spike, this feature is not possible dark matter halos due to intrinsic low volume densities and radially smooth density profiles. Figure 3 clearly demonstrates the need for the complete rotation curve to determine the true (physical) dynamic mass properties of any galactic disk.

It should not be alarming that the peak velocity ($V_{Esc}$) is twice $V_{Circ}$. For example, isothermal models have $V_{Esc}$=√2$V_{Circ}$, while dark matter halos (NFW) typically are in the range $V_{Esc}$≈2.5$V_{Circ}$ with a dependency on halo concentration (Cattaneo 2017).

*Evidence for Virial Dynamics within the Milky Way Disk*
The existence of a dynamic discontinuity at $R_P$ is revealed in extended probes of stellar counts in the halo and the number distribution of the Galaxy's dwarf galaxy satellites (Lopez-Corredoria 2018) (Kelley 2019), respectively. Figure 4 shows the observed number of stars LAMOST (blue) and SDSS-APOGEE (green) and the Galaxy's satellite count via Gaia DR2 (black). The virial radius $R_P$, (vertical red dot) and outer disk edge $R_D$ (vertical black dash).

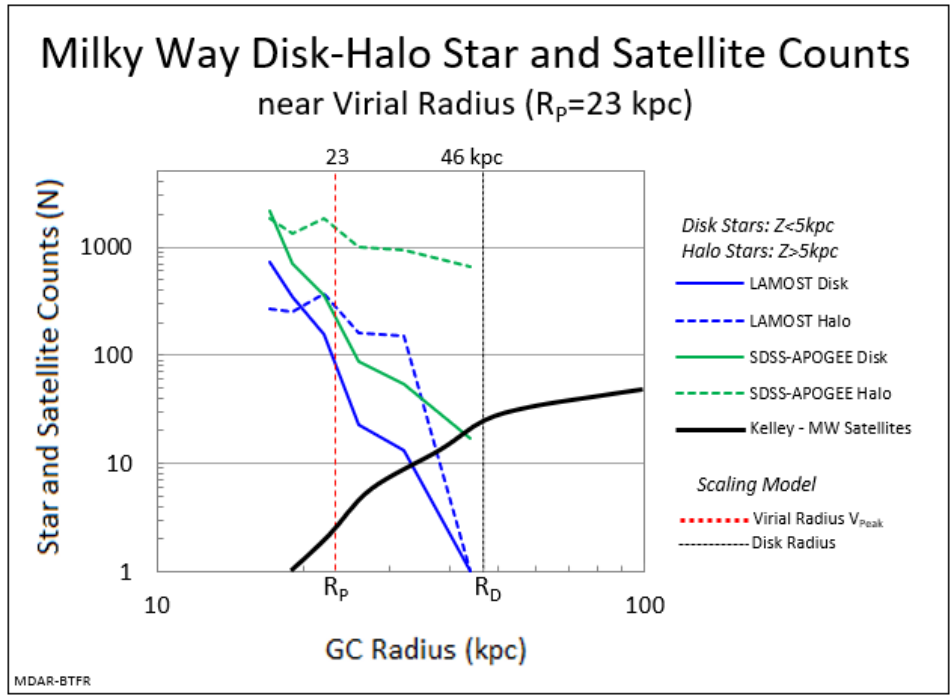

*Figure 4: Milky Way disk-halo star and satellite counts as a function of radial distance from the GC. Disk scaling parameters $R_P$ and $R_D$ are included for perspective. All curves are identified in the legend. The stellar disk extends to ~20 kpc followed by a severe outer truncation in star counts. Satellites demonstrate a corresponding inner truncation near the same radius. Image source – Tables 1 & 2 (Lopez-Corredoria 2018), Fig. 7 (Kelley 2019) (Bobylev 2017)*

The above figure illustrates the severe truncation of Galactic disk star counts for LAMOST and SDSS data sets in near $R_P$. Beyond $R_P$, the dearth of stars indicates a clearing of orbits close to the Galactic plane. The halo stars show a similar decline in counts but not to the degree that occurs in the disk. For satellites, the reverse is apparent where the truncation occurs inside $R_P$ - evidence for a significant break in dynamics near this specific radius. This same feature has been examined in great detail using Gaia and SDSS local main sequence as well as distant (halo) BHB kinematics (Deason 2018). Although this kinematic feature has been well documented, it has not been adequately addressed in the mainstream literature.

*Galactic Rotation Curves – a Scaling-Homology Perspective*
One of the more perplexing aspects of galactic rotation is the "unexpected diversity" now being observed in dwarf galaxy surveys (Oman 2015). In this paper, we simplify understanding of rotation curve diversity via Newtonian mechanics and its three scaling parameters; dynamic mass ($M_D$), disk radius ($R_D$) and average circular velocity ($V_C$). Not previously considered, are that galactic virial radii are located within $R_D$, not as extended as halo property $M_{200}$ (~200 to 300 kpc). For simplicity, the model fixes $R_P=0.5R_D$ providing 'first-order' geometric alignment and consistency for the Galactic dynamic.

To group physical and dynamical properties of six example galaxies shown in Table 1, the Ponomareva galaxy classification scheme is employed. Galaxies are binned into three categories: "rising" ($V_{max} < V_{flat}$) that is associated with dwarf galaxies, classically "flat" rotation ($V_{max} = V_{flat}$), and "declining" profiles ($V_{max} > V_{flat}$) that define early-type spirals (Ponomareva 2017). This is a much simpler method to categorize galaxies than based on their individual rotation curves.

Table 1 lists the physical scaling properties of six galaxies that span a wide range of mass, size and rotation velocities. The observed scaling values; $M_{Dyn}$, $R_{Disk}$ and $V_{Circ}$ are obtained from published rotation curves in the table footnote. The Ponomareva profile type for each galaxy is given in column 7 by comparing the rotation velocity at $R_P$ (=0.5$R_D$) versus $R_D$.

*Table 1: Scaling Parameters for Six Example Galaxies*

| Example Galaxies | | $M_{Dyn}$ | $R_{Disk}$ | $V_{Circ}$ | $V_{Peak}$ | Rotation | Surface Den.[b] |
|---|---|---|---|---|---|---|---|
| Name | Type | $\times 10^{12} M_\odot$ | kpc | kms$^{-1}$ | kms$^{-1}$ | Profile Type[a] | $\mu_{RD} = M_\odot pc^{-2}$ |
| DDO 64 | Im/BCD | 0.001 | 2.2 | 45 | 31 | $V_{Peak} < V_{Circ}$ | 66 |
| And IV | VLSB dIrr | 0.0034 | 7.5 | 45 | 45 | $V_{Peak} = V_{Circ}$ | 19 |
| IC 2574 | dIrr | 0.0145 | 13.5 | 68 | 68 | $V_{Peak} = V_{Circ}$ | 25 |
| F586-3 | LSB | 0.055 | 16.3 | 120 | 120 | $V_{Peak} = V_{Circ}$ | 66 |
| Milky Way | HSB | 0.50 | 40 | 230 | 432 | $V_{Peak} > V_{Circ}$ | 99 |
| M31 | HSB | 1.47 | 100 | 251 | 467 | $V_{Peak} > V_{Circ}$ | 47 |

And IV $\mathcal{D}_{Obs}$ = 9.1 (others $\mathcal{D} \equiv 5.9$)  [a] $V_{Peak}$ @ 0.5$R_{Disk}$  [b] $M_{Dyn}/\pi(R_{Disk})^2$

References: DDO 64 and F586-3 (Kuzio de Naray 2010), IC 2574 (Oman 2015), And IV (Karachentsev 2016), M31 (Sofue, Dark Halos of M31 and the Milky Way 2015), and the MW; $V_{Circ} \leq 40$ kpc (Pato 2015), $V_{Circ} > 40$ kpc (Sofue, Dark Halos of M31 and the Milky Way 2015), $V_{Obs}$ 6 ≤ $R$ ≤ 30 kpc (King III 2015) and $V_{Obs} < 4$ kpc (Reid 2014). Appendix A and B provides rotation curves used to determine the scaling parameters for DDO 64, And IV, IC 2574 and F586-3 (with $\mathcal{D}$=5.9 at $R_D$ equivalent to baryonic fraction $f_b$=0.17, agreeing with the ΛCDM cosmological mean value).

The dynamic surface densities ($\mu_{RD}$) for each galaxy will prove crucial in modeling the observed galactic acceleration profiles and the RAR. This parameter plays an important role in the $g_{Bar}$-$g_{Obs}$ (acceleration) relationship observed in the SPARC dataset provided by McGaugh, Lelli and Schombert (McGaugh 2016).

We have purposely selected three galaxies (And IV, IC 2574 and F586-3) having similar circular velocities at $R_P$ and $R_D$ - classicaly 'flat' rotation ($V_{Peak} = V_{Circ}$). Mass discrepancies are fixed at $\mathcal{D}$ =5.9 except for galaxy And IV which is an observed value.

In Figure 5, we compare the six galaxies in the conventional 'R-V' (linear) format illustrating rotation velocity as a function of radius. The colored symbols represent observed "scaling solutions" for each galaxy obtained from their individual rotation curves and cut-off at $R_D$ using Table 1 convention. Most galaxies cluster along the [$M_{Dyn}$-R-V] relation (purple curve) with most variance due to dynamic surface density values. In the figure below, the [$M_{Dyn}$-R-V] relation is shown for illustrative purposes. It provides a reference for subsequent figures and discussion.

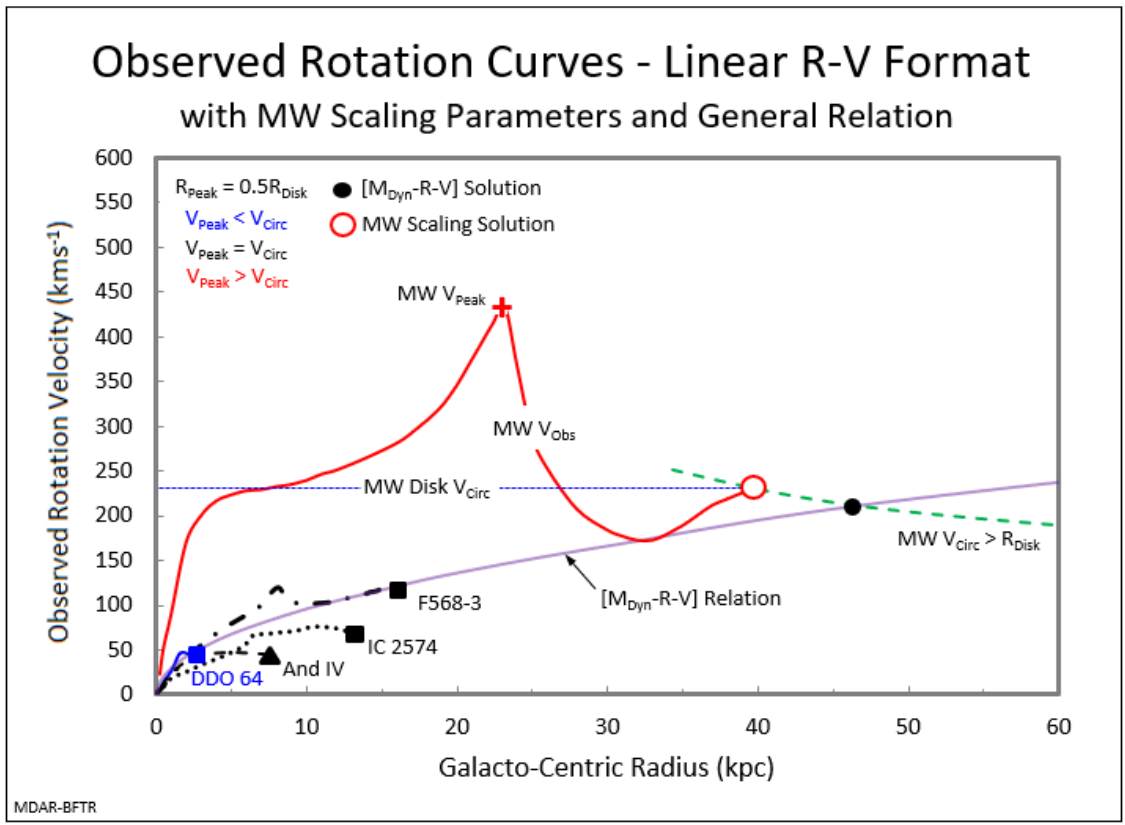

*Figure 5: The conventional linear R-V format with five galactic examples, including Ponomareva rotation curve categorization. MW scaling parameters include; rotation curve (red solid), $V_{Circ}$ (blue dash), Keplerian decline > $R_D$ (green dash), and $V_P$ (red-cross) and the "scaling solution" (open red circle). Disks tend to cluster near the "cosmological" R-V mean value (purple solid) termed the [$M_{Dyn}$-R-V] relation. The black point located on the Keplerian curve represents the R-V scaling solution ($R_D$=48.6 kpc and $V_{Circ}$=210.2 kms$^{-1}$) for a galaxy with a mean surface density $\mu_{[MRV]}$=67$M_\odot$pc$^{-2}$. Galaxies with R-V scaling solutions above [$M_{Dyn}$-R-V] have dynamic surface mass greater than $\mu_{[MRV]}$ while those below have lower surface density. M31 is not included as its disk radius ($R_D$) is 100 kpc, well beyond the range depicted in the figure. M31's disk length is highly extended (for its mass) due to relatively low surface density - roughly one-half of the MW.*

The linear-linear 'R-V' format above graphically demonstrates the huge diversity between galaxies but its form is not appropriate for power-law analysis and revealing physical scaling relations between galactic parameters. To this end, most figures presented in the balance of the paper are presented as log-log and normalized in terms of dimensionless scaling ratios.

Figure 6 below depicts five galaxies re-scaled and normalized with $R_D$=1 and $V_{Circ}$=1 at $R_D$. All disks intersect at 1:1 ($V_C$, $R_D$). In addition, all galaxies demonstrate a conventional Keplerian declines (green dash) beyond $R_D$. This unique format reveals disk features and scaling relations not possible with a linear 'R-V' format. For each galaxy, mid-point ($R_P$) velocities are identified (crosses) for comparasion to the circular velocity at $R_D$.

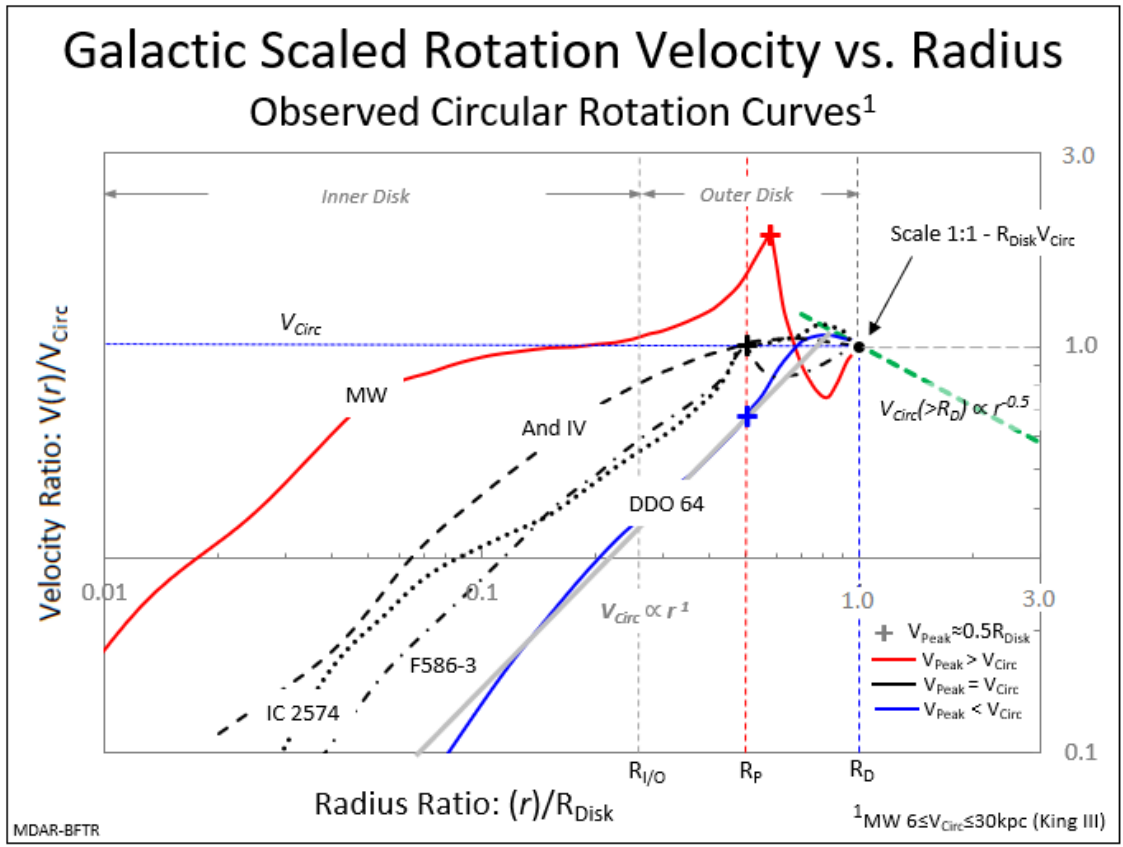

*Figure 6: Dimensionless log-log R-V scaling format for DDO 64, IC2574, F586-3, And IV, and the MW. Most disk inner regions demonstrate near-rigid body rotation with DDO64 having rigid body rotation to ~$R_D$ (gray solid). "Intermediate" galactic rotation velocities are variable but intersect $V_C$=1 at 0.5$R_D$ (black cross) corresponding to Ponomareva class $V_{Peak}$=$V_{Circ}$. Per the scaling model, all disks obey Keplerian mechanics beyond $R_D$ (green dash) in accordance with Newtonian mechanics.*

Employing the dimensionless log-log format shown above, the galactic curves exhibit several interesting features:

- All galaxies exhibit varying degrees of differential rotation. DDO64 demonstrates rigid body characteristics nearly throughout its disk.
- Outer disk rotation profiles are substantially more diverse than those in the inner disk. Each of the three Ponomareva "intermediate" galaxies (black curves) exhibit rotation curves that intersect at $R_P$ at $V_{Circ}$. These rotation curves intersect at 0.5$R_D$, each via a different kinematic "path."
- F568-3 exhibits a small velocity peak at 0.5$R_D$, which should exceed ~200kms$^{-1}$. If this peak is greater than the stated circular velocity, it would shift this galaxy's classification to $V_P$>$V_C$.
- Escape velocity profiles are dependent on galactic dynamic distribution (surface density) and total mass showing substantial deviation from the Newtonian expectation ($V_{Esc}$ = $\sqrt{2}V_{Circ}$).
- The scaling expectation beyond $R_D$ are expressed as the classical Keplerian decline, consistent with central point mass with $M_{Dyn}$=$R_D V_C^2$/G. No galaxy should be considered an "outlier."

By rescaling galactic disk parameters based on observed properties at the galactic edge ($V_C$, $R_D = 1$), a true differentiation between rotation curves is possible (Figure 5). Rather than attempting to achieve universal models, the focus should be on the variation and direct causes for it (Oman 2015).

*The [$M_{Dyn}$-R-V] Relation for Rotationally Supported Galaxies*

In accordance with Newtonian dynamics and the virial theorem, all disk galaxies should exhibit consistently robust scaling relationships. Much work has been performed in determining these relations for simple spherically-shaped dispersion supported early-type galaxies. The observationally derived relation shows $M_{Dyn}$ is proportional to $R_c\sigma^2$, a formula comparable to Newton's law (Cortes 2017). A similar relation has been found to describe rotationally supported galaxies as well. For these galaxies, the analogous relation is $M_{Dyn} \propto R_{Disk}V_{Circ}^2$ which is simply Newton's law for circular motion (La Fortune 2016a).

Figure 7 below represents a deprojected log-log three-dimensional parameter space for $R_{Disk}$, $V_{Circ}$ and $M_{Dyn}$ with six example galaxies positioned per their individual physical parameters. The MW galaxy (open red circle) also includes its virial velocity peak (red cross) associated with $M_{Dyn}=1.0 \times 10^{12} M_\odot$ at $R_P=23$ kpc. And IV and DDO64 galaxies are highlighted (horizontal blue dot) as having the same circular velocity. Note that their separation in the plot describes two very different types of galaxies. For perspective, a representative MW-sized dark matter halo (gray/black circle) is included.

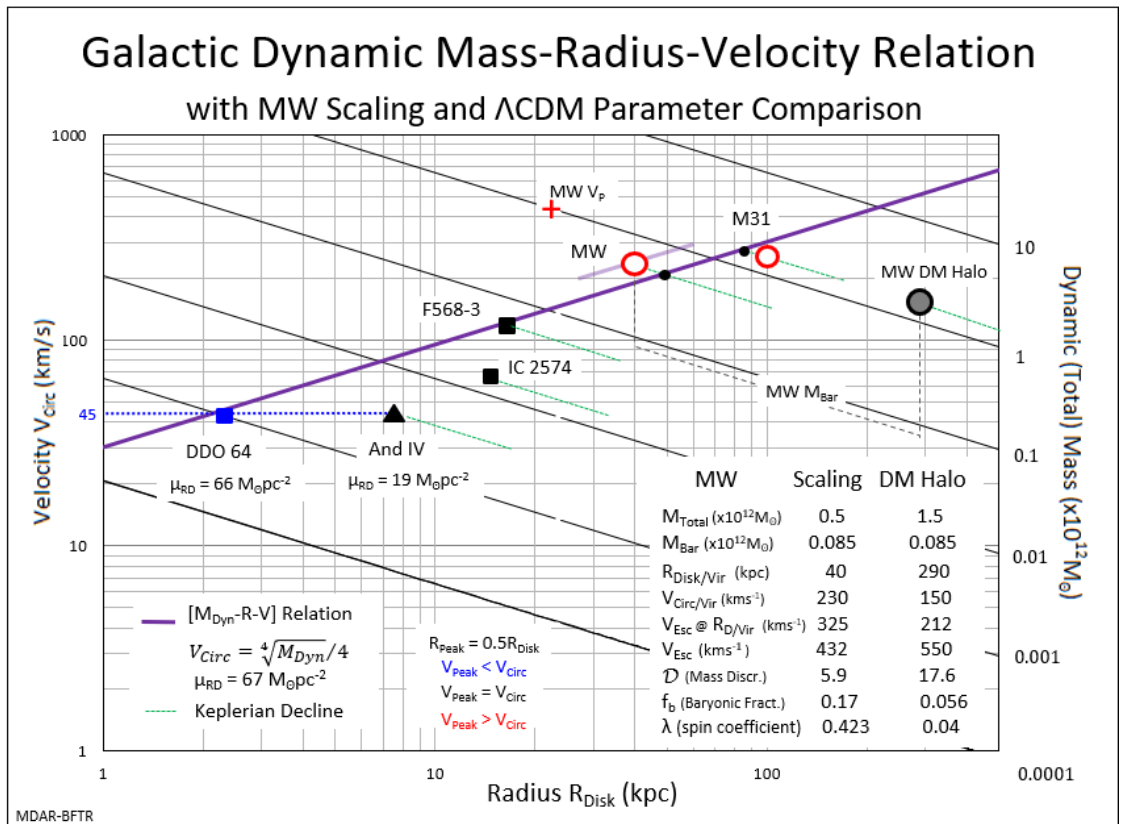

*Figure 7: A log-log R-V plot and the deprojected $M_{Dyn}$. The [$M_{Dyn}$-R-V] relation (purple dash) indicates the central tendency for galaxies to fall along this universal curve. Dynamic mass surface densities follow along the [$M_{Dyn}$-R-V] for a wide range of $M_{Dyn}$. For comparison, a MW-sized dark matter halo is positioned in the grid (gray solid circle) with all physical properties listed in the inset. Thin black dash lines*

*representing the Milky Way's $M_{Bar}$ content links the scaling solution to the dark matter halo indicating that the halo greater mass discrepancy than the scaling model. Galaxy color-coding is per Ponomareva classification with galaxy parameter listed in Table 1. In the lower left, the reduced, dimensionless form of the [$M_{Dyn}$-R-V] relation is shown.*

In the above figure, a 'median' dynamic mass surface density is evident and identified as the [$M_{Dyn}$-R-V] relation (solid purple). This particular solution reduces to a simple analytical equation between circular velocity and dynamic mass in a form analogous to the Tully-Fisher relation. Units used for paper are $M_{Dyn}$ [$M_\odot$], $V_{Circ}$ [kms$^{-1}$], and $R_D$ [kpc]:

$$V_{Circ} = \sqrt[4]{M_{Dyn}}/4$$

Per Newton's law of circular motion, the corresponding disk radius is obtained:

$$R_{Disk} = 16G\sqrt{M_{Dyn}}$$

Note that the above two equations are only accurate solutions for galaxies on the [$M_{Dyn}$-R-V] relation. To account for the wide range of galactic dynamic surface densities, a simple ratio is used between observed $\mu_{RD}$ [$M_\odot$pc$^{-2}$] and $\mu_{[MRV]}$ = 67$M_\odot$pc$^{-2}$, the [$M_{Dyn}$-R-V] value:

$$V_{Circ} = \sqrt[4]{(\mu_{RD}/\mu_{\{MRV\}})\,M_{Dyn}}/4$$

$$R_{Disk} = 16G\sqrt{(\mu_{[MRV]}/\mu_{RD})\,M_{Dyn}}$$

All three parameters in the above equations are interrelated, maintaining relational consistency between parameters. For example, as $\mu_{RD}$ decreases, so does $V_{Circ}$ in maintain a constant dynamic mass $M_{Dyn}$. Therefore, $R_{Disk}$ likewise must increases in complementary fashion. This analysis can be appreciating that the dimensional unit for the gravitational constant $G$ is acceleration divided by mass surface density (Christodoulou 2018).

*The Physical Basis for the Baryonic Tully-Fisher Relation (BTFR)*
While galactic disks show substantial variation in disk dynamic surface density, all are governed on a fundamental level by the universal [$M_{Dyn}$-R-V] scaling relation. The second equation is the familiar Tully-Fisher relation for galactic dynamic mass (substituted for $M_{Bar}$):

$$V_{Circ} = \sqrt[4]{M_{Dyn}}/4 \text{ or } M_{Dyn} = 256V_{Circ}^4,$$

The Baryonic TFR is obtained through the mass discrepancy ratio $M_{Bar}$=$M_{Dyn}/\mathcal{D}$ with $\mathcal{D}$=5.9 as our fiduciary value. This is the scaling version of the BFTR:

$$M_{Bar} = \left(\frac{256}{5.9}\right)V_{Circ}^4 = 43V_{Circ}^4$$

The scaling zero-point value agrees well with the empirical value $M_{Bar}$ = 47±6$V^4_{flat}$ (McGaugh 2012). The range around the measured zero point is equivalent to a relatively narrow span of mass discrepancy $\mathcal{D}$=4.8 to 6.2.

*The Origin of the Radial Acceleration Relation (RAR)*

In this section, we investigate the Radial Acceleration Relation with application of the formally employed Mass Discrepancy Acceleration Relation (MDAR). A thorough analysis of the Spitzer Photometry & Accurate Rotation Curve (SPARC) dataset has been completed and it has been found that the RAR fit describes the acceleration data very well using $a_0$ as the only free-parameter and a specific interpolating function (McGaugh 2016). Figure 8 provides their results with SPARC-based accelerations shown as binned averages (red squares). The empirical RAR (black solid) is obtained from the data. Added to the original figure are two mass discrepancy iso-contours; $\mathcal{D}$ = 5.9 (green dash) and 12.1 (black long dash). One point often overlooked is that all data is contained within the region spanning $\mathcal{D}$= 1 to 12.1 for any measured $g_{obs}$.

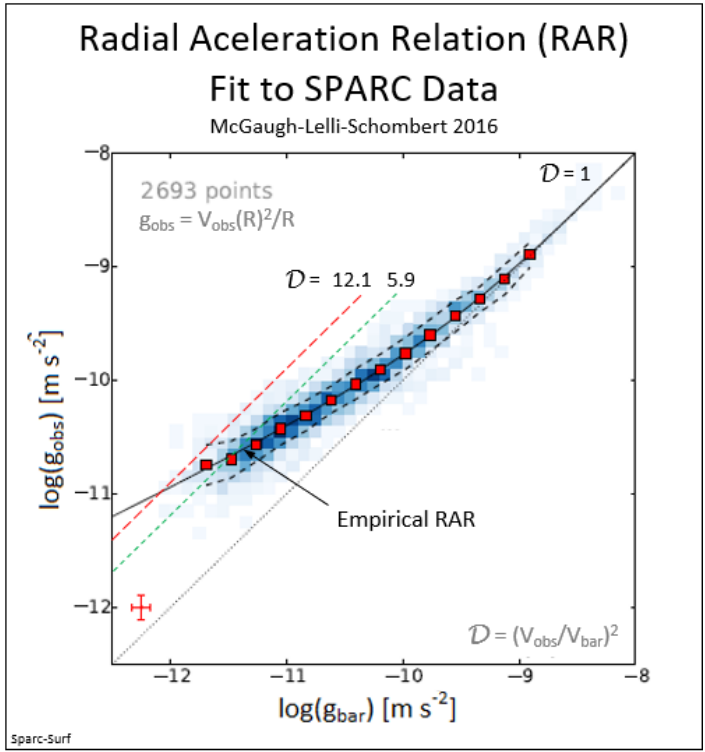

*Figure 8: The McGaugh-Lelli-Schombert (MLS) SPARC data and resultant Radial Acceleration Relation (black solid). Mass discrepancy accelerations for $\mathcal{D}$=5.9 (green dash) and $\mathcal{D}$=12.1 (black dash) are included for reference for following figures. Image source – Fig. 3 (McGaugh 2016)*

The authors described the RAR as obtained by the SPARC dataset using a single parameter, $g_†$ and a value equal to Milgrom's constant $a_0$=1.2x10$^{-10}$ms$^{-2}$.

$$g_{McG} = \frac{g_{bar}}{1 - e^{-\sqrt{g_{bar}/g_†}}} \quad \text{with} \quad g_† = a_0$$

The RAR has advantage over competing acceleration/rotation curve fits in that it has only one free parameter that well describes a wide variety of galaxies. MOND adherents perceive the constancy of $a_0$ and overall simplicity of its "form and fit" as evidence of a new natural law attributable to disk galaxies (Lelli 2017) (McGaugh 2014). We show that this empirically derived new law is a manifestation of conventional mechanics applied to a self-gravitating, distributed matter disk.

In Figure 9, we reproduce the McGaugh-Lelli-Schombert RAR (gray solid) and add the positions of the six example galaxies based on mass discrepancy and measured terminal disk accelerations. Mass discrepancy iso-contours and past findings by Burkert and Gentile (black circle with error bars) are included for reference. The Burkert/Gentile studies indicated dark matter halo, independent of galaxy type, all attain a maximal universal or 'characteristic' acceleration, log $g_{dark}(r_0)$ ≈ -10.5 and a 'characteristic' baryonic contribution log $g_{bar}(r_0)$ ≈ -11.24 (ms$^{-2}$) as well corresponding to $\mathcal{D}$=5.6. The scaling results for our six galaxies agree with this general conclusion, but with variability due to the impact of dynamic mass surface density differences between disks.

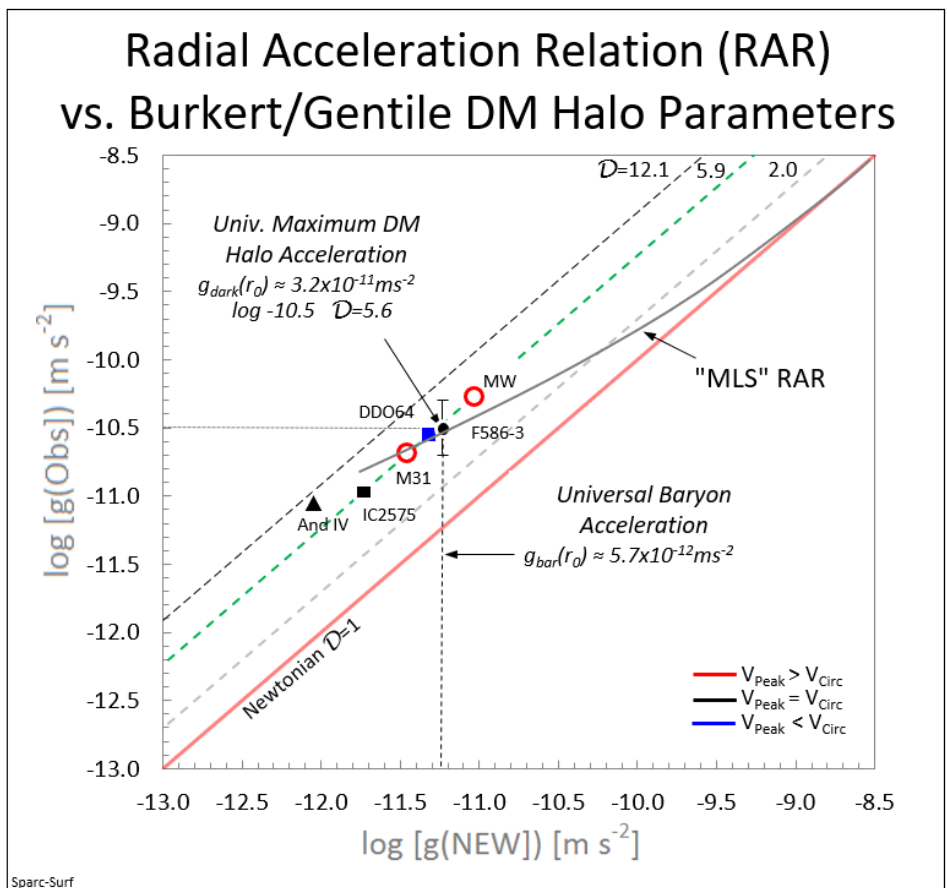

*Figure 9: Scaling modeled acceleration data for six example galaxies (see Ponomareva key) versus the Newtonian $g_{bar}$ expectation (red solid). Mass discrepancies for 2, 5.9, and 12.1 are shown parallel to the Newtonian curve (various colored dash). The MLS RAR (gray solid) intersects Burkert/Gentile observed universal maximum DM halo acceleration (black circle) indicating a strong agreement between past and recent studies. Except for And IV ($\mathcal{D}$ =9.1 obs.), the remaining five galaxies were modeled with $\mathcal{D}$=5.9 (green dash).*

A key to understanding this relation is the location of each galaxy compared to the empirical RAR (gray solid). Not all galaxies trace the RAR, except those having dynamic surface densities near $\mu_{RD}$=67 M$_\odot$pc$^{-2}$ as are DDO64 (blue square) and F586-3 (black square, partially hidden). From Figure 9, it is evident that dynamic mass surface density impacts the length and positioning of the RAR within the acceleration space. We illustrate the relation between galactic dynamic surface density and disk acceleration in Figure 10 below. The Burkert/Gentile result (black circle / vertical black dash) is compared against the [$M_{Dyn}$-R-V] solution (vertical purple solid).

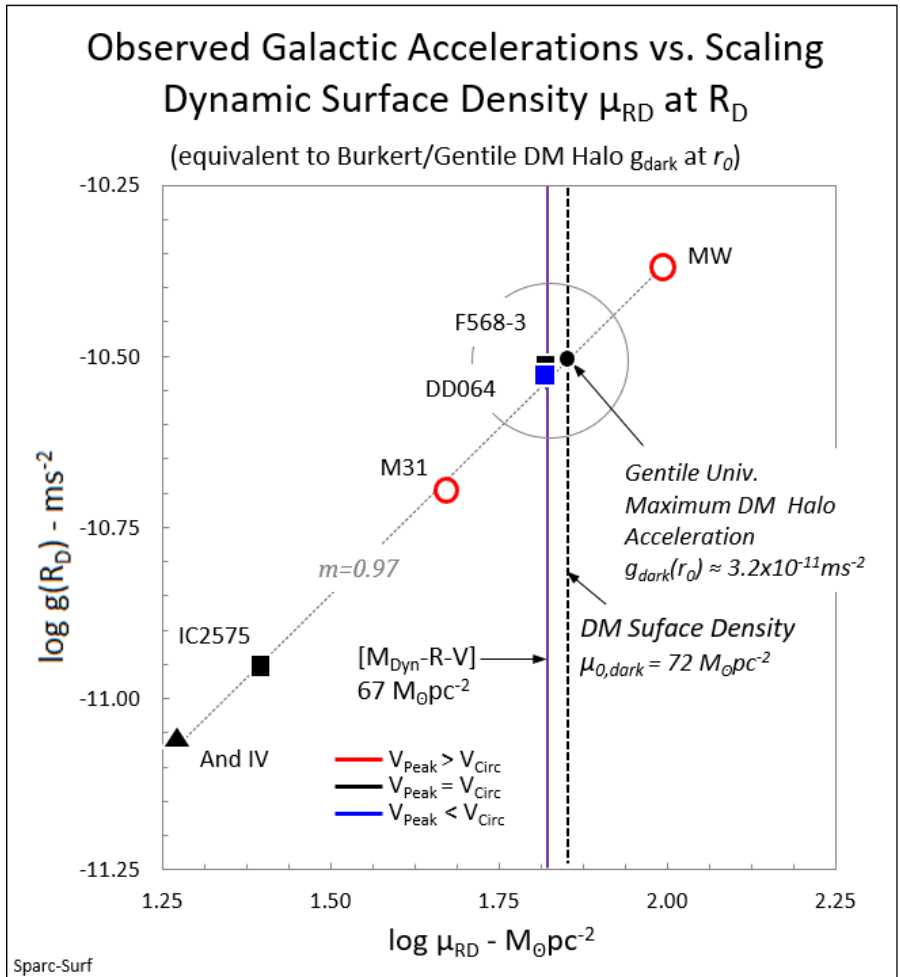

*Figure 10: Comparison between Burkert/Gentile and scaling results. In addition to confirming previous findings, we uncover a direct correspondence between terminal disk acceleration and total galactic dynamic surface density. This dependency has not been previously investigated, but is evident in the tight correlation between parameters.*

As expected, there is close agreement between the Burkert/Gentile maximum DM halo acceleration at $r_0$ and the value obtained at $R_D$ via the scaling method. Using dynamic surface density as a parameter, we link two observables, velocity and radius to galactic accelerations and the solution for the RAR with respect to $V_{Circ}$ (for any radius inside $R_D$). This is the first analytic expression of the RAR in terms of purely physical properties:

$$g_{obs} = V_{obs}^2/r = \pi G \mu_{(r)} \rightarrow V_{Obs} = \sqrt{\pi G \mu_{(r)} r}$$

It has been previously stated that the RAR is a "natural consequence" of the BTFR (Wheeler 2019). The scaling model confirms this "consequence" provided all physical constraints are respected. In Table 2, we illustrate the inherent scalability of this approach within galactic disks for several important parameters.

Table 2: Newtonian[1] Galactic Disk Scaling Relations

| r | $\mu(r)$ | $g_{bar}(r)$ | $g_{obs}(r)$ | $M_{obs}(r)$ | $\rho_0$ | $r^2\rho_0$ |
|---|---|---|---|---|---|---|
| 1 | 1 | 1 | 1 | 1 | 1 | 1 |
| 0.5 | 2 | 4 | 2 | 0.5 | 4 | 1 |
| 0.25 | 4 | 16 | 4 | 0.25 | 16 | 1 |
| 0.125 | 8 | 64 | 8 | 0.125 | 64 | 1 |
| 0.0625 | 16 | 256 | 16 | 0.0625 | 256 | 1 |
| 0.03125 | 32 | 1024 | 32 | 0.03125 | 1024 | 1 |
| $R_D=1$ | $\mu(R_D)=1$ | $g_{bar}(R_D)=1$ | $g_{obs}(R_D)=1$ | $M_{obs}(R_D)=1$ | $M_{obs}/r^3$ | $M_{obs}/r$ |

[1] $M_{Dyn} = R_D V_{Circ}^2/G$ for $r \leq R_D$

The expected linear $M_{Dyn} \propto R$ relation inside $R_D$ is confirmed employing four recently published independent estimates (Eadie 2018) (Gnacinski 2018) (Posti 2019) (Watkins 2019). This 1:1 relationship is illustrated in the Figure 16 (inset) found in Appendix D. It is not surprising that the scaling solution is identical to the Burkert halo model with $r^2\rho_0$ = constant.

The above scaling relations make possible an analytic solution to the empirically derived Radial Acceleration Relation (RAR) (McGaugh 2016). As in accord with MONDian tenets we define the scaling RAR as the acceleration required to support a constant (or flat) circular velocity throughout the disk. In Figure 11, the physical RAR is supported by two acceleration contributions, a steeper $M_{Bar}$ acceleration (red solid), and a shallower distributed dynamic sourced from the disk dynamic (blue solid). The resultant combination results in the scaling RAR (black solid). This particular RAR curve applies to any galaxy existing on the [$M_{Dyn}$-R-V] relation with $\mu_{RD}=67 M_\odot pc^{-2}$. This is only one set of all potential RAR curves physically possible. For comparison, the empirical RAR by McGaugh, Lelli, and Schombert is included (gray solid).

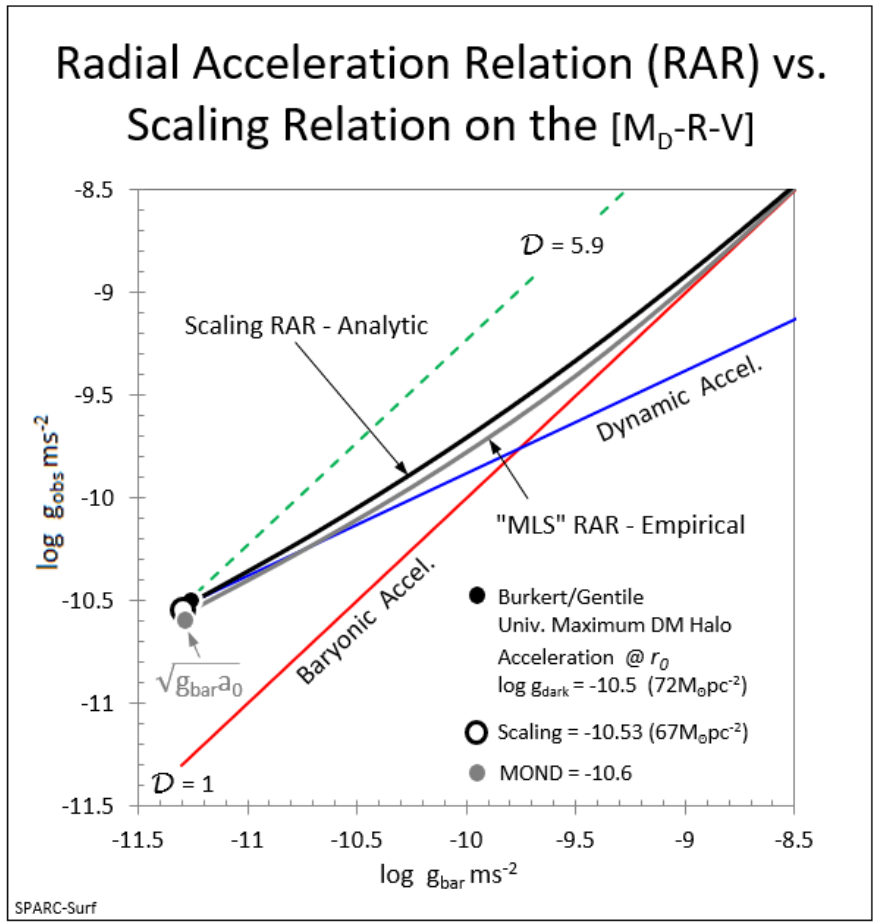

*Figure 11: Scaling RAR (black solid) denoting acceleration to maintain constant $V_{Circ}$ throughout the disk. SPARC derived empirical RAR is shown (gray solid) and estimated $g_{obs}$ at $R_D$. Burkert/Gentile halo universal accelerations (black dot) is consistent with scaling value (open black circle) and MOND value. The scaling RAR comprises acceleration components, one baryonic (red solid) and the other dynamic (blue solid).*

In Figure 11, the scaling RAR traces physical disk acceleration required to maintain a flat/constant circular velocity throughout the disk. With the empirical RAR, a slight downward displacement is observed compared to the scaling expectation. This difference may be due to the SPARC sample averaging a lower galactic dynamic surface density than the [$M_D$-R-V] valued and/or variation from the scaling mean mass discrepancy $\mathcal{D}$=5.9 (green dash). No universal scaling equation is provided as each galaxy's morphological and dynamical peculiarities modify the function. Most importantly, unlike MOND-like solutions, the scaling RAR does not require a universal acceleration parameter. It appears that the scaling RAR effectively maps the combined baryonic and dynamic accelerations and implies disks obey Newton's law for circular motion with proper accounting of disk energy and angular momentum. There is very close agreement between the scaling and empirical RARs as well as the calculated MOND acceleration (gray point) determined at the disk's edge via $\sqrt{g_{bar}a_0}$. MOND's popularity is not that is fundamental, but that its simple form faithfully describes physical scaling relations as exists in the galactic setting.

*Summary*

We propose an alternative, physical-based approach to galactic dynamics based on conventional law and empirical physical relations. These relations account for intrinsic disk related energy and angular momentum. This approach offers a physical solution to the 'missing mass' problem linked to ongoing thermodynamics and a reliance on the gravitational content of ordered motion/potential energy (Bruschi 2016). Perhaps a motivating force for observed 'non-Newtonian' disk dynamics can be traced back to relativistic effects during galactic formation in a rapidly expanding universe (Telkamp 2018). It is not suggested that this model can explain all unresolved problems, but that a fresh look at the 'missing mass' problem from a classical perspective is warranted.

*Acknowledgement*

We thank anonymous referees for assisting in development of this proposal, personal guidance and feedback.

*Appendix A – DDO 64 and Andromeda IV Observed Rotation Curves/Scaling Parameters*

Figure 12 shows galactic rotation curves and fits for DDO64 (ΛCDM) and And IV (isothermal) are shown for reference purposes.

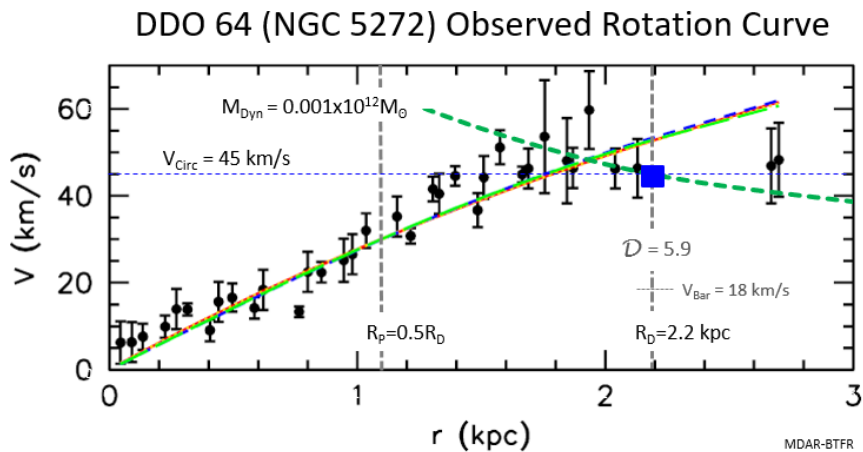

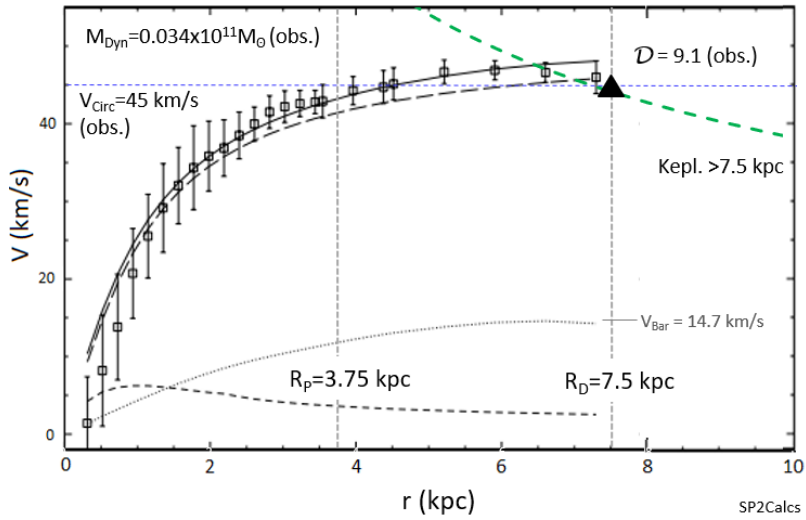

*Figure 12: Observed galactic rotation curves for DDO 64 – Fig. 1 (Kuzio de Naray 2010) and Andromeda IV – Fig. 8 (Karachentsev 2016). The Keplerian declines (green dash) beyond $R_D$ provides a measure of total dynamic mass. The intersection between $V_{Circ}$ at $R_D$ associated consistent with dynamic mass determines the scaling parameters found in Table 1. Per toy scaling model, $R_P=0.5R_D$.*

### Appendix B – F-568-3 and IC 2574 Observed Rotation Curves/Scaling Parameters

Figure 13 shows galactic rotation curves with ΛCDM fits for F586-3 and IC2574 are shown for reference purposes.

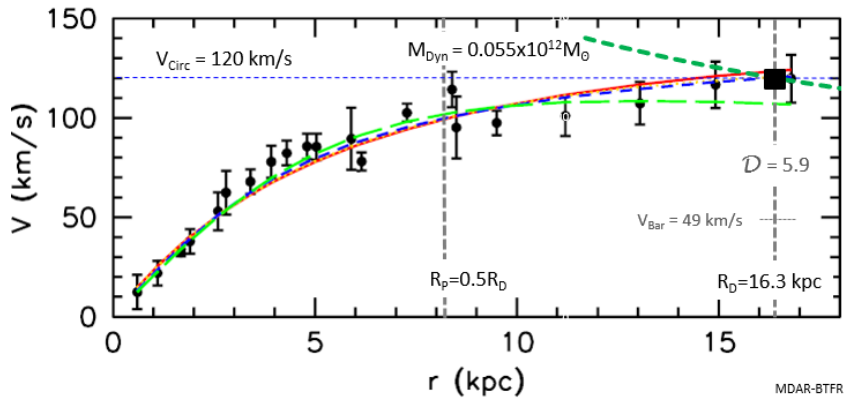

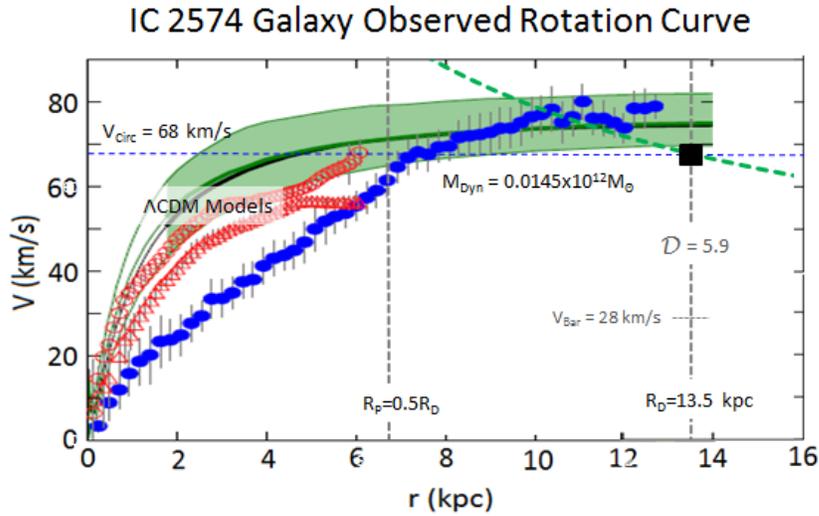

*Figure 13: Observed galactic rotation curves for F568-3 – Fig. 1 (Kuzio de Naray 2010) and IC 2574 – Fig. 1 (Oman 2015). Scaling parameters are included for completeness. The Keplerian declines (green dash) beyond $R_D$ provides a measure of total dynamic mass. The intersection between $V_{Circ}$ at $R_D$ associated consistent with dynamic mass determines the scaling parameters found in Table 1. Per toy scaling model, $R_P=0.5R_D$.*

### Appendix C – Observed Galactic Stellar Halo Velocity Dispersion Components

King's observed component velocity dispersion results (R, θ, and φ,) are shown below in Figure 14 (King III 2015). Included in the original are the isotropic component dispersions (black dash) corresponding to a central mass $M_{Bar}=0.085 \times 10^{12} M_\odot$ and Galactic circular velocity $V_{Circ}=230 \text{ kms}^{-1}$ (blue dash). The lower right-hand panel provides the anisotropy coefficient β, indicating highly tangential, high velocity orbits in a narrow region of the halo/disk near $R_P=23$ kpc.

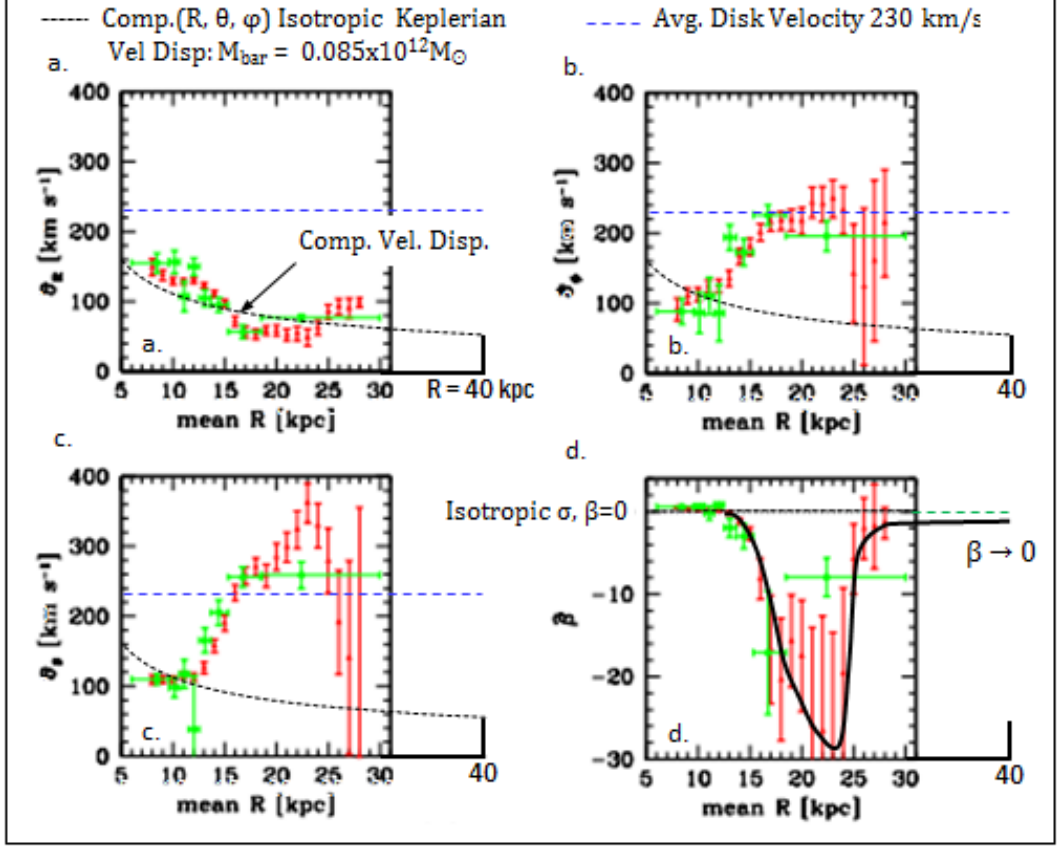

*Figure 14: King III observed dispersion components (red / green) as a function of Galactic radius (King III 2015). Point mass isotropic component dispersions corresponding to a central mass $M_{Bar}=0.085 \times 10^{12} M_\odot$ (black dash) and Galactic circular velocity $V_{Circ}=230 kms^{-1}$ (blue dash) are indicated in panels a., b., and c. Panel d. provides the radial function of the anisotropy coefficient β indicating highly tangential velocities in a narrow region of the halo/disk.*

In the above figure, the radial dynamic mass distribution within the Galaxy is visualized as tracing the net positive difference in observed velocity and the baryonic isotropic curve (black dash). The quadrature sum of all three components at 23 kpc is $V_{Peak}$=432 kms$^{-1}$ equivalent to the Maxwell-Boltzmann most probable velocity $V_{MP}$. This interpretation is also supported by kinematics suggesting a virial origin for the Hyper Velocity Star population (La Fortune 2016b).

*Appendix D – Milky Way Dynamic Mass Profiles to 400 kpc: Scaling-homology and ΛCDM Models vs. Observation*

Observations to 400 kpc clearly reveal the thermodynamic nature and dynamic mass profile of the Galaxy. This is illustrated in Figure 15 for data compiled from several sources over this wide expanse (Fig.1) (Bajkova 2017). In this figure, Bajkova's "best fit" ΛCDM Model III (black solid) is provided for comparison to the scaling parameters. Model III's mass profile smoothly and monotonically increases with radius and asymptotically flattens to $M_{DM}=0.75 \times 10^{12} M_\odot$ (light gray dash). Overall, ΛCDM Model III sufficiently fits the data, but offers no more than a general trend with a lone data point inside $R_{Disk}$.

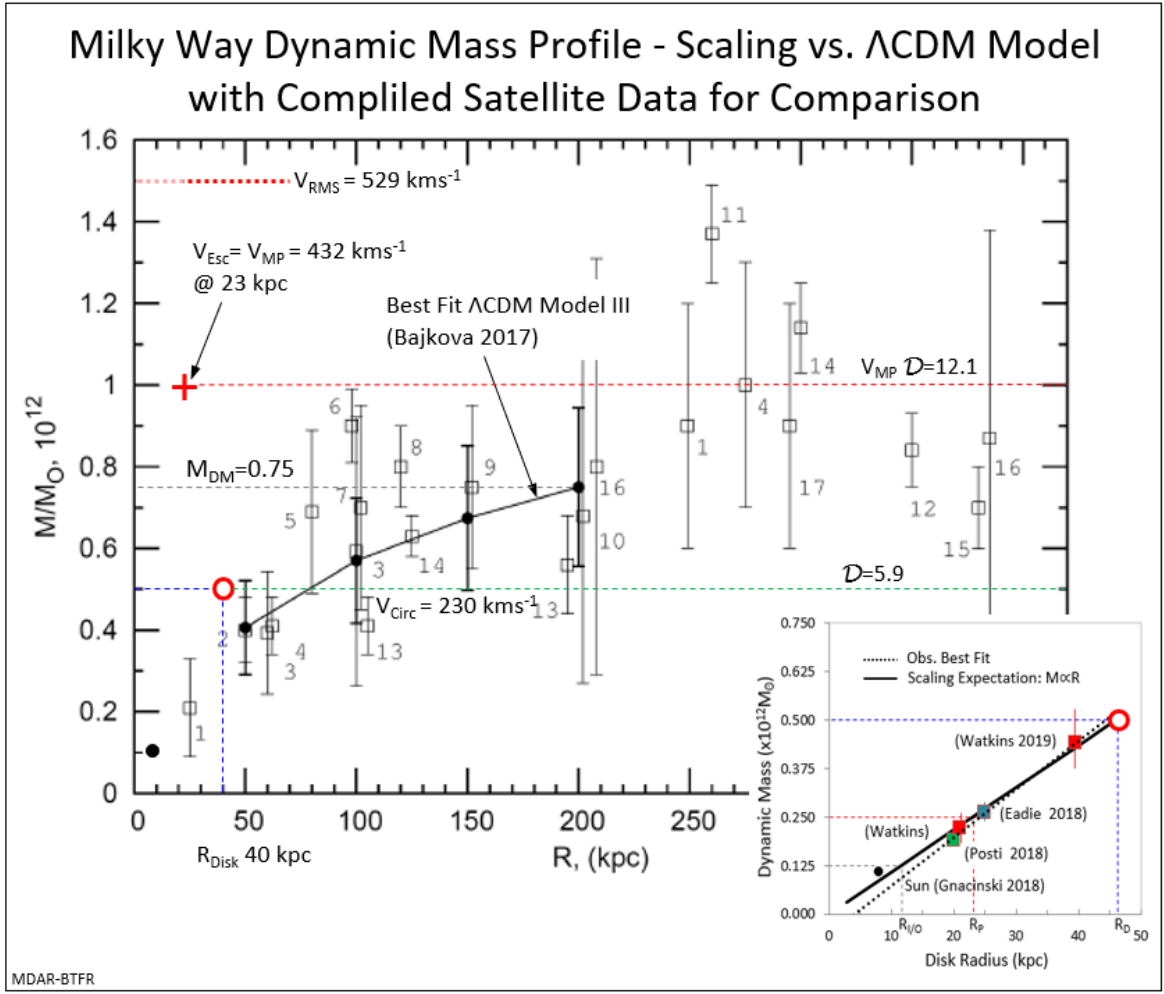

*Figure 15: Large plot: complied satellite data expressed as Galactic dynamic mass as a function of radius (Bajkova 2017). Bajkova best fit ΛCDM model III (solid black). Observed scaling parameters included for reference. Inset: MW dynamic mass estimates inside disk with an observed mass/radius ratio ≈1. In both images, the scaling "solution" is the intersection of $V_{Circ}$ and $R_{Disk}$ (red open circles).*

Per above, the scaling solution at $V_{Circ}$-$R_{Disk}$ (red circle) is associated with a dynamic mass $M_{Dyn}=0.5 \times 10^{12} M_\odot$ equivalent to $\mathcal{D}=5.9$. Likewise, the virial mass is $M_{Dyn}=1.0 \times 10^{12} M_\odot$ ($\mathcal{D}=12.1$) and is associated with the Maxwell-Boltzmann most probable velocity $V_{MP}=432$ kms$^{-1}$ at 23 kpc. (red cross). The M-B distribution extends to greater velocities with $V_{RMS} \approx 1.5 \times 10^{12} M_\odot$ (red dotted). Of course, any orbiting body will experience this mass outside the virial radius. It is not surprising that ΛCDM Model III's mass splits the difference between the scaling virial and circular dynamic mass estimates. In fact, we find Model III mass is the Newtonian escape velocity value √2$V_{Circ}$ equal to $\mathcal{D}=8.34$ for $M_{Dyn}=0.5 \times 10^{12} M_\odot$.

It is interesting to note that an early examination of the Galactic rotation curve to 400kpc using ΛCDM cosmological models provided a mass $M_{halo}=1.04 \times 10^{12} M_\odot$ (Sofue 2011). A more recent estimate based a similar model and *6-D* Gaia data is $M_{Tot}=1.00 \times 10^{12} M_\odot$ (+0.31-0.24) (Deason 2019). As interpreted in the scaling model, this coincides with virial mass. It is precisely double the disk mass as derived as from Newtonian circular motion with a flat $V_{Circ}$. Determining precise Galactic dynamic mass estimates is not due to a lack of data but of interpretation. In fact, Deason estimates total mass discrepancy at $\mathcal{D}$=11.1. This value is associated with the thermodynamic threshold $\mathcal{D}$=12.1 as determined from observed peak dispersion velocity at the Galactic virial radius. The $V_{RMS}$ value is similar to larger MW dark matter halos in the range 1.2 to $1.6 \times 10^{12} M_\odot$. The wide velocity dispersion inherient in the Maxwell-Boltzmann probability distribution is responsible for this uncertainty in Galactic total mass.

Recently an update has been made to Deason's estimate, increasing total mass to $M_{Tot}=1.29 \times 10^{12} M_\odot$ (Grand 2019). This updated mass estimate is between 1.0 and $1.5 \times 10^{12} M_\odot$ as associated with values of $V_{RMS}$ and $V_{RMS}$, respectively. Grand's mass till maintains a factor of two ($0.82$-$1.66 \times 10^{12} M_\odot$) "uncertainty" reflective of the wide range of velocities present in the Galaxy. The M-B probability distribution hampers efforts to obtain an unambiguous precise mass. This issue is best illustrated in a comprehensive summary of Galactic mass estimates (Wang 2019). As illustrated in this compilation, one can reasonably consider a M-B velocity distribution as responsible for the disparate mass estimates, not modeling artifacts or errors in the methods used. With the advent of Gaia DR2, there is adequate data to physically interpret and understand observed Galaxy kinematics.

In the Figure 15 inset, several dynamic mass estimates are reported <$R_D$. Combining three independent studies, the expected $M_{Dyn}$:$R_D \approx 1$ ratio is confirmed (Eadie 2018) (Gnacinski 2018) (Posti 2019) (Watkins 2019). A simple linear fit (black dotted) agrees with the 1:1 scaling expectation (black solid. The dynamic and virial masses are two separate entities and must be treated as such rather than lumping all dynamics into a simplistic dark matter halo and providing an "intermediate" fit to the general data set.

*Appendix E – The Scaling Milky Way Dynamic Mass Toy Model and Comparisons with Gaia DR2 High Velocity Star (HVS) Observations*

A toy scaling model is described and built to investigate the radial dynamic mass profile of the Milky Way only using baryonically associated energy ($E$) and angular momentum ($J$) per the Peebles spin equation as exists in the disk setting (λ=0.432). The model utilizes a small number of conservative parameters and employs the true rotation velocity ($V_{Obs}$) as a function of disk radius, including all velocity components. This velocity includes support from rotation and dispersion components that encodes information regarding the nature and interrelationships governing $E$ and $J$ contributions along the observed rotation curve. This scaling approach reverses conventional modeling methodology. Rather than "matching" $V_{Obs}$ to arbitrary (ad hoc) models by invocation of external agents such as dark matter or non-Newtonian dynamics, we rely on internal or intrinsic velocity support associated with the baryonic disk. Individual $E$ and $J$ values are obtained via the spin parameter. The components of the toy model, including model assumptions, governing relations and Galactic parameters are:
$M_{Dyn}=0.5 \times 10^{12} M_\odot$, $V_{Circ}=216$ kms$^{-1}$ and $R_{Disk}=46$ kpc, dynamically consistent with virial velocity $V_{Peak}=432$ kms$^{-1}$ at $R_{Peak}=23$ kpc. We model the baryonic mass as a log normal distribution (Marr 2015a):

$$M_{Bar}(r) = \frac{\Sigma_0}{r/r_\mu} \exp\left(-\frac{[ln(r/r_\mu)]^2}{2\sigma^2}\right) \text{ with } \Sigma_0 = 0.07, r_\mu = \ln(R_D), \sigma = 1.68 \text{ for } r > 0$$

This particular distribution describes an isolated, collisionless ensemble of gravitationally self-interacting particles having maximal entropy at $R_P$. In Figure 17 below, for disk radii greater than 3 kpc, the log normal distribution is represented as a super-position of co-spatial baryonic exponentials and other (cored gas disk) distributions, each having their individual respective central surface densities and scale lengths.

The log normal baryon distribution parameters ($\Sigma_0$, $r_\mu$, and $\sigma$) are derived by simultaneously solving two interrelated observational constraints. Inside $R_P$, Galactic dynamic mass ($M_{Dyn}$=0.5x10$^{12}$M$_\odot$) is consistent with the Keplerian escape velocity ($V_{Esc}(R_D)$= √2$V_C$= 305 kms$^{-1}$) *and* $\mathcal{D}$ = ($M_{Dyn}$/$M_{Bar}$) = 5.9. The second constraint is $V_P(R_P)$=432 kms$^{-1}$ *and* $\mathcal{D}$ = 12.1.

Under the above two constraints, the complete $V_{Obs}$ curve and the log normal $M_{Bar}(r)$ distribution are used to calculate E as governed by the classical integral of motion (Kipper 2016):

$$I_1 = v_R^2 + v_\theta^2 + v_\phi^2 - 2\Phi \text{ with } \Phi = \text{Potential Energy and } V_{Obs}^2 = v_R^2 + v_\theta^2 + v_\phi^2$$

Using the general equation, E=1/2mv$^2$, the Galactic energy profile $E(r)$ can be plotted as a function of disk radius including dispersion components through $V_{Circ}$= √($v_R^2 + v_\theta^2 + v_\phi^2$):

With:

$$E(r) = \frac{M_{Bar}(r)V_{Circ}^2(r)}{2}$$

Similar to E, the disk's radial angular momentum (J) profile is obtained using the rotational and azimuthal components of $V_{Obs}$ and the integral of motion:

$$I_2 = Rv_\phi \text{ with } v_\phi^2 = V_{Circ}^2 + \sigma_\phi^2 \text{ (azimuthal vel. disp.)}$$

The angular momentum profile $J(r)$ of the disk follows the general equation as:

$$J(r) = r\, M_{Bar}(r) V_\phi(r)$$

The combined E and J velocity support is fixed by the Peebles spin parameter equation in the disk setting:

$$\lambda = 0.423 = \frac{J\sqrt{E}}{GM_{Dyn}^{5/2}}$$

For the log normal baryon distribution (and *G*=1) the overall dynamic mass can be expressed in terms of the E and J components as a function of disk radius:

$$M_{Dyn}(r) = \sqrt{2} J(r)^{2/5} E(r)^{1/5} \qquad (1/0.423^{2/5} \cong \sqrt{2})$$

For this example, we map the observed rotation curve inside the virial radius ($R_P$). Note that this equation holds for the entire disk, but that the E and J components are not well defined between $R_P$ and $R_D$ due to significant kinematic disruption.

The results of the toy model are presented below in Figure 16. Baryons and associated E and J contributions are plotted as a function of disk radius in a log-log format 'normalized' to coordinates $R_{Disk}$=1 and $V_{Circ}$=1. The red open circle at ($V_{Circ}$, $R_D$) represents the MW's physical scaling "solution" that parameterizes the global dynamic. The red-cross is positioned at ($R_P$, $V_P$) in the outer disk equivalent to a localized effective virial mass $M_{Dyn}=1.0\times10^{12}M_\odot$, twice the global $M_{Dyn}$ value at $0.5R_D$. All disk components are identified in the key.

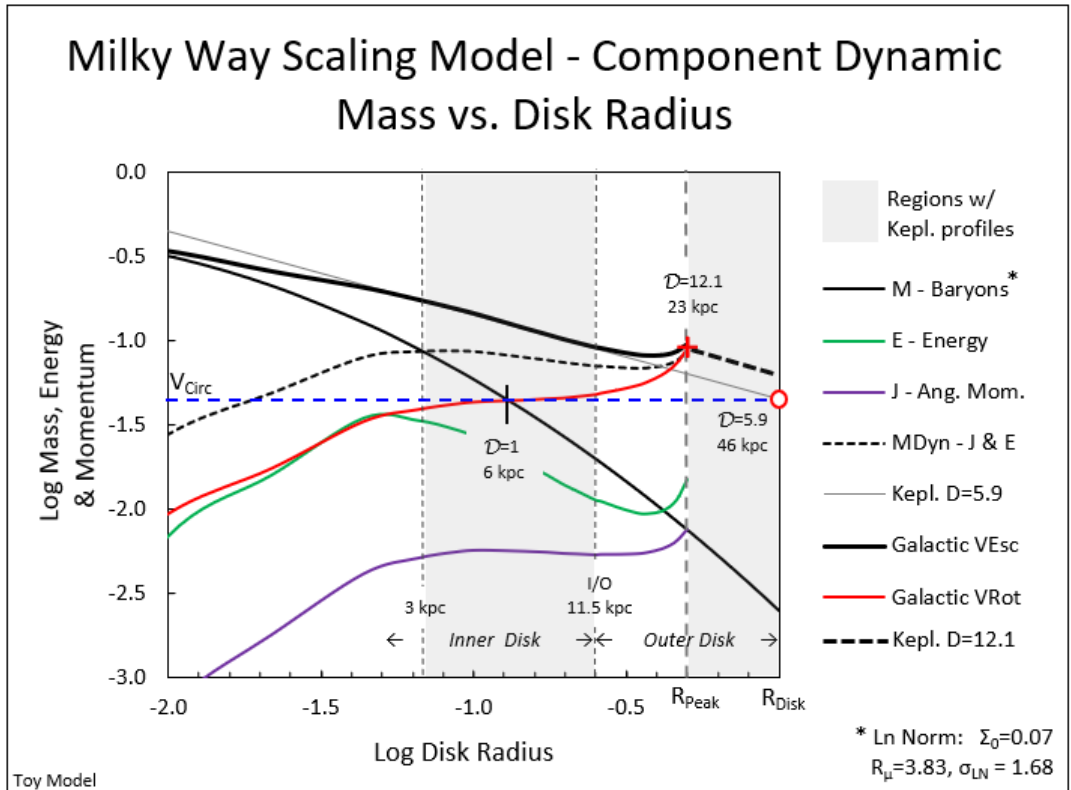

*Figure 16: Toy scaling model dynamic components (identified in key) as a function of disk radius. In particular, the shaded radial regions are Keplerian in nature. The peak dynamic mass at the virial radius $R_{Peak}$ ($\mathcal{D} = 12.1$) is dynamically equivalent to the local velocity $V_{MP}$. The Galactic E and J disk contributions to the overall circular velocity is relationally quantified. The global dynamic is represented by the upper-most curves (black solid to $R_P$ and black dash from $R_P$ to $R_D$).*

In above figure, two radial regions are highlighted (gray shade). Per the model, these regions exhibit (1/√r) kinematics with total mass tracing a conventional Keplerian dynamic and this expectation is tested using recently observed High Velocity Star (HVS) observations from Gaia precision proper motion data. For this comparison, we employ the findings of two published DR2 surveys, both targeting the high velocity cohort via a minimum velocity cut (>~450 kms$^{-1}$) (Marchetti 2018) (Hattori 2018). Per scaling, these stars represent the population residing on or near the global virial 'surface' that fixes local $V_{Esc}$. In addition to anticipated Keplerian radial profile, the predicted velocity distribution should quantitatively describe a Maxwell-Boltzmann characteristic as well.

Figure 17 below is drawn from Marchetti (Fig. 2) representing the entire data set including selected high velocity star candidates. Focusing on the inner region, yellow triangles identify possibly bound stars while red circles identify possibly unbound stars. Light blue circles reflect bound stars having very precise proper motions helping demarcate the physical Galactic escape velocity "envelope" within the inner disk. Completing the figure is the virial peak, $V_P$ at $R_P$ (red-cross) and the scaling solution, $V_{Circ}$ at $R_D$ (open red circle). As predicted in the toy model in Figure 16, the two shaded regions illustrate simple Keplerian dynamics with $V_{MP}$ and $V_{RMS}$ defining the R-V space inhabited by the HVS population.

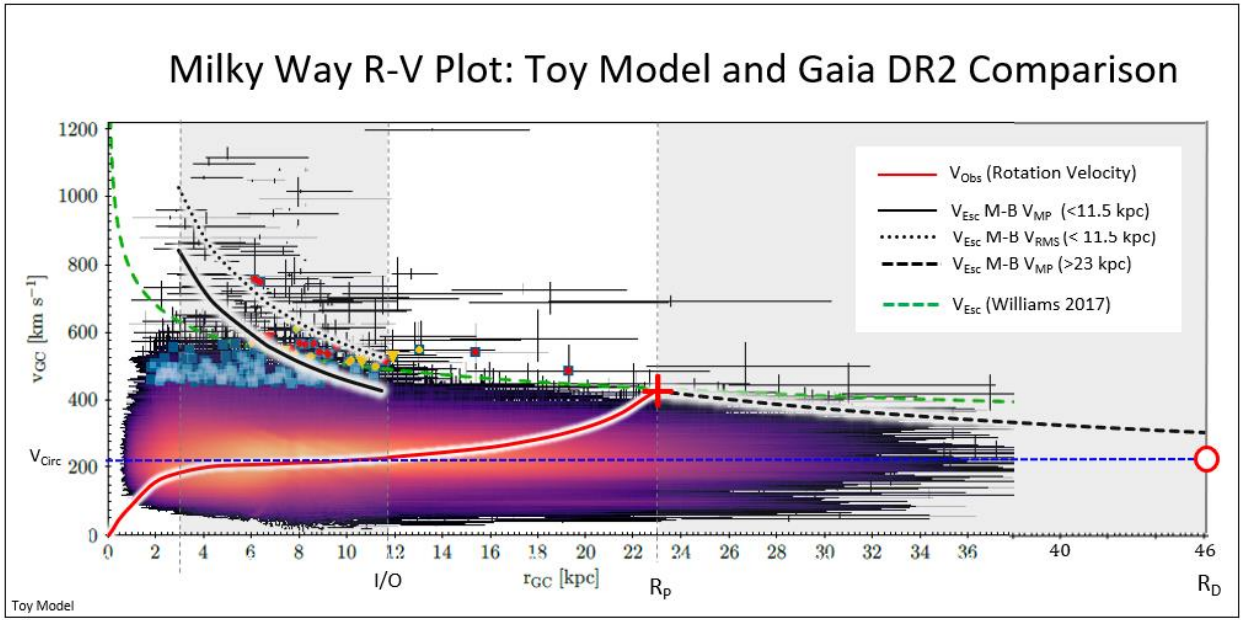

*Figure 17: Gaia DR2 velocity/radius data including selected HVS candidates (red circles and squares, yellow circles and triangles) (Marchetti 2018). Williams estimated Galactic escape velocity (green dash) against toy scaling model illustrated in Figure 17. Shaded regions denote Keplerian dynamics in the inner (black solid - $V_{MP}$, black dotted - $V_{RMS}$) and outer disks (black dashed - $V_{MP}$). The HVS sample reflects expected scaling dynamics to very high precision employing the latest proper motion data. The blue HVS data is considered mostly bound.*

Above, within the inner disk (left-hand shaded region) we express the regime of stars near or at escape velocity in terms of a Maxwell-Boltzmann distribution with $V_{MP}$ (black solid) and $V_{RMS}$ (black dotted) following the Keplerian dynamic as a function of radius. The overall kinematics are both Keplerian and statistical in nature. The outer (right-hand shaded region) also depicts a Keplerian decline as traced by Gaia DR2 data high velocity cohort. The central (unshaded region between $R_{I/O}$ and $R_P$) exhibits a near constant escape velocity suggesting 'pseudo-isothermal' dynamics in the inner-mid disk region. It is expected that $V_{Esc}$ throughout this region is fairly independent with radius. The Williams $V_{Esc}$ curve (green dash) has been retained from the original Marchetti figure.

The origin of the HVS population has been a longstanding controversy. The most popular theory is that the majority of these fast stars should originate close to the Galactic Center (GC). This, as a result of gravitational black hole "ejection" due extreme gravity conditions required to achieve such high velocities. With Gaia DR2 proper motion data, it is now possible to extract reliable "source directions" from individual star radial and tangential velocity components. This is best illustrated using a "Toomre' Diagram" as shown in Figure 18 below (Marchetti Fig 3).

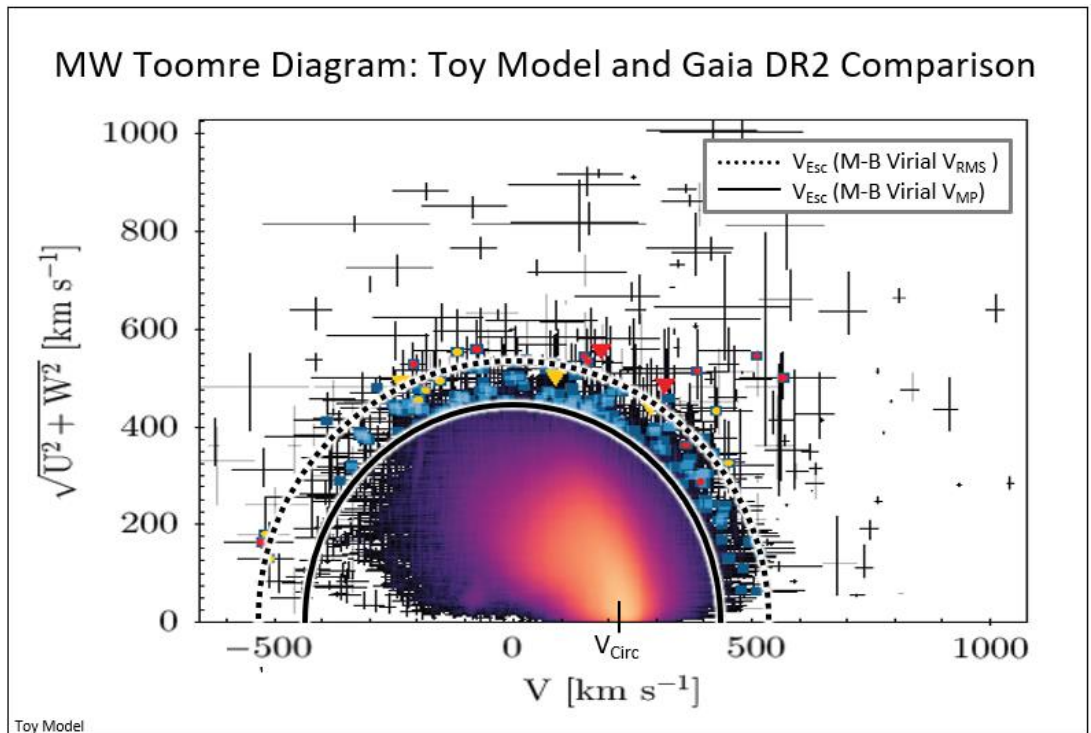

*Figure 18: Toomre' diagram of the Marchetti HVS sample. Loosely bound stars are identified as red and yellow data points* (Marchetti 2018). *Note that all are bounded from below by $V_{MP}$ (black solid) and roughly above by $V_{RMS}$ (black dotted) in a nearly uniform fashion spanning a wide range of orbital properties and angular momentum.*

The Toomre' format provides information regarding the potential origin (location) through proper motions. If black hole "ejection" mechanisms are the source of HVS, then Gaia data should cluster near zero angular momentum (near the $V_{MP}$ boundary at V≈0). Upon general inspection, we find most tighter bound HVS (light blue squares) cluster near $V_{Circ}$ indicative of a disk origin as these still carry significant angular momentum. More importantly, the majority of these fast stars are centered on the Galactic rest frame, occupying a well-defined region between $V_{MP}$ and $V_{RMS}$ (black solid and black dotted, respectively). Rather than central black hole ejection, we find a more distributed population drawing into question central ejection as the primary source. Rather than rare chance encounters, the HVS population represents evidence for a thermodynamic interpretation.

Similar to Marchetti, Hattori also investigated "extreme velocity stars" inside 8 kpc using Gaia DR2 (Hattori 2018). His Fig. 2 is depicted in Figure 19 below. To the original Hattori figure, the Keplerian velocity profile is included with the $V_{MP}$ and $V_{RMS}$ radial profiles shown.

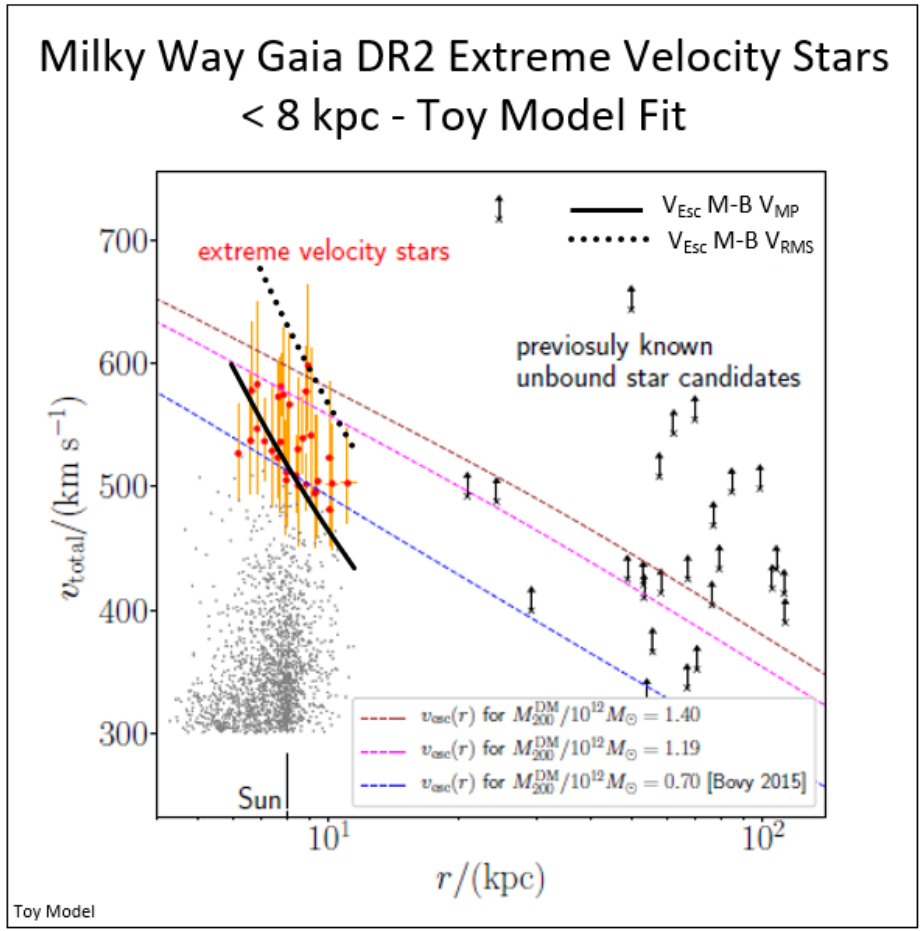

*Figure 19: Extreme Velocity Star velocity/radius data from Gaia DR2 (Hattori 2018). As with preceding figures, local scaling escape velocities are shown (black solid – $V_{MP}$, black dotted – $V_{RMS}$). The Keplerian dynamic inside 10 kpc (< $R_{I/O}$) is evident with highest velocity stars situated within the expected $V_{Esc}$ range dictated by global disk dynamics and total mass. Note that the statistically-based M-B velocity dispersion is a source of the large uncertainty in MW total mass estimates being between 0.7 and $1.4 \times 10^{12} M_\odot$ shown in the reference figure (blue, magenta, and brown dash).*

In the above figure, Hattori identified ~30 extreme velocity stars (red data points with yellow error bars) from the dataset. Out of this sample, they found five that *may* have originated from Galactic center or the Large Magellanic Cloud. Based on kinematics, the rest of the sample has no definitive source or origin identified. Hattori concluded ejection mechanisms do not provide the complete picture regarding the origin of the HVS population.

In Figure 19, we see excellent agreement with these velocities falling between the Maxwell-Boltzmann $V_{MP}$ and $V_{RMS}$ curves indicating these stars are very near or at escape velocity. The $V_{MP}$ value at the solar radius (8 kpc) is equal to the escape velocity $V_{Esc}$= 521 kms$^{-1}$. This expectation is consistent with *6-D* Gaia observations which found $v_{esc}(r_0)$=528 kms$^{-1}$ at $r_0$=8.3 kpc (Deason 2019). The scaling value is within the error margins (±24) in Deason's analysis, but also provides a physical model of the escape velocity distribution.

*Appendix F – A Scaling Dynamic Mass Toy Model of M31 with Globular Cluster Counts and Kinematics*

The toy model reflects a universal template for massive spiral galaxy dynamics. In this section we construct a homologic facsimile for M31 with $V_P=2V_{Circ}=V_{Esc}$ and $R_P=0.5R_D$ with $\mathcal{D}=12.1$ at $R_P$. The M31 model is an "exact copy" of the Galaxy's, only differing in scale. Based on rotation curve analysis, scaling parameters for M31 are $V_{Circ}=233$ kpc and $R_D=116$ kpc. A significant difference is the dynamic surface density of M31 $\mu_{RD}=34 M_\odot pc^{-2}$ which is less than half of Milky Way. This relatively low surface density explains why M31's $V_{Peak}$ is only 35 kms$^{-1}$ greater than the Milky Way, although triple the dynamic mass.

In Figure 20, the observed Globular Cluster (GC) population of M31 is plotted as a function of their individual radial velocity and distance (Veljanoski 2013) (Fig.3). The work is similar to Lopez-Corredoria radially binned star counts which exhibited a sharp truncation in stellar density at the Milky Way's virial radius $R_P=23$ kpc. As with the Milky Way's stellar halo component, M31 demonstrates a corresponding truncation in the globular cluster sample at its virial radius $R_P=58$ kpc. As with the Milky Way example, we highlight (gray shade) two radial regions exhibiting simple Keplerian escape velocity profiles. The globular cluster data is compiled from two surveys: RBC – gray with error bars near the core and Veljanoski data (squares with error bars) located beyond 18 kpc. Three streams are identified by color as shown in the key.

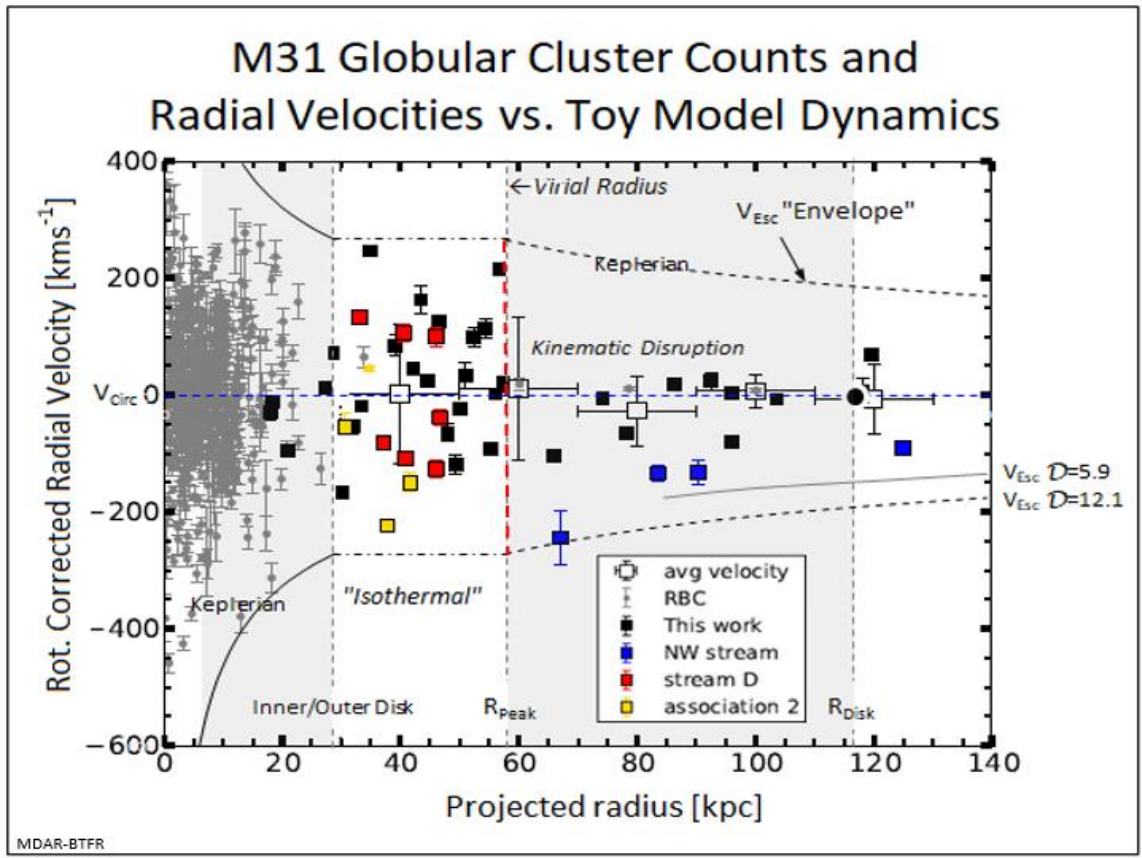

*Figure 20: M31 Globular Cluster population as a function of radial velocity and projected distance from center (data identified in key) (Veljanoski 2013). Highlighted are the toy model parameters for M31 which include the two Keplerian regions (gray shade) as with the MW. There is a significant break in dynamics at $R_{Peak}$ due to virial processes. This "surface" boundary splits the quasi-isothermal region inside $R_{Peak}$ and*

*the colder, isotropic dynamic outside. The NW stream (blue squares) appears to achieve local negative radial velocity $V_{MP}$ (467 kms$^{-1}$) near $R_P$. The NW stream is located inside the $V_{Esc}$ envelope (black dash) and is bound to M31 with high probability.*

In the above figure, we break the disk/halo into: inner/outer disk boundary, radius of virial (peak) velocity and total disk radius (all vertical gray dash) and M31's average circular velocity (horizontal blue dash). The virial radius (vertical red dash) is shown in as in toy model shown in Figure 2. As with the Milky Way example, we highlight (gray shade) two radial regions exhibiting simple Keplerian escape velocity profiles. The global escape velocity "envelope" of M31 is similar to the Milky Way, except for scale.

Apparent are radially segregated globular cluster radial distributions and velocities. It had been concluded that the inner RBC population density exhibits a defined break at 30 kpc, very close to the scaling derived 29 kpc inner/outer boundary (0.25$R_D$). Our interest lies with Veljanoski's data that captures a holistic portrait of M31's disk/halo potential. Focusing on the "pseudo-isothermal" region, we find the cluster velocity distribution extremely uniform with a net radial velocity close to zero. Just beyond the virial radius, $R_P$=58 kpc, the GC population count drops drastically. The outer halo clusters appear to be at apocenter and/or in highly circular orbits that may or may not coincide with the galactic plane of the disk.

The Veljanoski data provides a dynamic "snap shot" of the North West Stream (blue squares) entering the gravitational potential of M31 with the closest cluster nearing virial escape velocity $V_{Esc}$=467 kms$^{-1}$ (dotted black $\mathcal{D}$=12.1). Trailing clusters are streaming inward following the same trajectory and are expected to accelerate in similar fashion. The global dynamic does not allow captured disk/halo clusters to easily escape inside $R_P$ due to a rising potential.

As with the (Milky Way), M31 exhibits a severe truncation in (star), globular cluster counts indicative of a thermodynamic "boundary" demarcated by a pseudo-isothermal dynamic inside $R_P$ while experiencing severe disruption beyond (i.e., the dearth of GCs from 58 to ~80 kpc indicates a regime of instability). The fact that both galaxies share the same dynamic profile greatly diminishes the probability that unrelated, independent, stochastic dark matter halo accretion events are responsible. It is predicted that the radial magnitude and shape of the $V_{Esc}$ envelope is a feature common to nearly all massive spiral galaxies.